\documentstyle[amssymb,prc,aps,epsfig]{revtex}

\begin{document}
\title{Reconstruction of Hadronization Stage in Pb+Pb Collisions at 158A GeV/c }
\author{S.V. Akkelin$^{1}$, P. Braun-Munzinger$^{2}$, and Yu.M. Sinyukov$^{1}$}
\maketitle

\begin{abstract}
Recent data on hadron multiplicities in central Pb+Pb collisions at 158A
GeV/c at mid-rapidity are analyzed within the concept of chemical
freeze-out. A non-uniformity of the baryon chemical potential along the beam
axis is taken into account. An approximate analytical solution of the
hydrodynamic equations for a chemically frozen Boltzmann-like gas is found.
The Cauchy conditions for hydrodynamic evolution of the hadron resonance gas
are fixed at the thermal freeze-out hypersurface from analysis of
one-particle momentum spectra and HBT correlations. The proper time of
chemical freeze-out and physical conditions at the hadronization stage, such
as energy density and averaged transverse velocity, are found. \newline
\end{abstract}

\begin{center}
{\small {\it $^{1}$ Bogolyubov Institute for Theoretical Physics, Kiev
03143, Metrologichna 14b,Ukraine. \\[0pt]
$^{2}$ Gesellschaft f\"{u}r Schwerionenforschung, D-64291 Darmstadt, Germany.%
\\[0pt]
}}

PACS: 24.10.Nz, 24.10.Pa, 25.75.-q, 25.75.Gz, 25.75.Ld.

Keywords: relativistic heavy ion collisions, hadronization, chemical
freeze-out, kinetic (thermal) freeze-out, hydrodynamic evolution, hadron
resonance gas, quark-gluon plasma, inclusive spectra, HBT correlations.

Corresponding author: Yu.M.Sinyukov, Bogolyubov Institute for Theoretical
Physics, Kiev 03143, Metrologichna 14b,Ukraine. E-mail: sinyukov@gluk.org,
tel. +380-44-2273189, tel./fax: 380-44-4339492
\end{center}

\section{Introduction}

Systems which are created in ultra-relativistic A+A collisions at SPS and
RHIC are initially very dense and display collective behavior. The
quasi-macroscopic nature of those systems allows to study new states of
matter at the phase boundary between a hadron gas (HG) and the quark-gluon
plasma (QGP). Among the different methods to study it, an important role
belongs to precise measurement of hadronic observables. One- and
multi-particle hadron spectra, however, do not carry direct information
about the phase transition since one should expect strong interactions in
the hadron resonance gas after hadronization. Usually, different theoretical
approaches are used to diagnose the matter properties at the transition
point. A rather fruitful concept is that of chemical freeze-out. It implies,
first, that the initial hadronic state is in chemical equilibrium \cite{P.
Br.-M.-1,Stachel,P. Br.-M.-2}. Second, the chemical concentrations of HG do
not change during the evolution. The latter assumption is justified by the
observation that the rate of expansion in HG is larger than the rate of
inelastic reactions and less than that of elastic collisions \cite{Koch}.

It gives the possibility to use the approach of local thermal equilibrium
and hydrodynamic expansion of HG at approximately frozen chemical
composition. The initial conditions for the hadronic stage of the collective
expansion are determined by matter evolution in the pre-hadronic phase and
also by particle interactions and hadronization conditions.

If chemical freeze-out takes place, it is defined by a temperature and
chemical potentials conjugated to conserved quantum numbers. To extract
those thermodynamic parameters from total particle number ratios one has to
assume uniform temperature, baryon and strangeness chemical potentials as
well as a unique hadronization hypersurface $\sigma ^{ch}(x)$ for all
particles. The result is a common effective volume for all hadron species,
that completely absorbs effects of flow and a form of the hypersurface and
approximately cancels in a description of particle ratios \cite
{Sinyukov,Cleymans}.

In detail, the situation may be somewhat more complicated. While baryon
stopping seems to be fairly complete at AGS energies \cite{E-AGS} (see also
Ref.\cite{Shengqin} where possible deviations from the full stopping are
analyzed quantitatively), there are clear indications for an onset of
transparency at top SPS energy from the observed fairly flat rapidity
distribution of pions in Pb+Pb central collisions \cite{NA49-1,NA49-4} and,
in particular, from the difference in rapidity distributions between protons
and antiprotons, which cannot be explained by longitudinal flow developing
after full stopping. As a consequence, one could expect particle number
ratios taken in mid-rapidity, especially involving nucleons, to differ
somewhat from those integrated over $4\pi $. On the other hand, a consistent
hydrodynamic analysis then must be based on an analysis of data in a narrow
mid-rapidity region, which is done in this paper. We also note that, in a
boost-invariant scenario as observed at RHIC, chemical freeze-out conditions
are the same for any individual rapidity slices taken within a wide enough
rapidity interval.

The object of this paper is to reconstruct the hadronization stage within
the hydrodynamic approach. We assume that hadronic matter at that stage has
locally a maximal entropy. Effects of hadron-hadron interaction (excluded
volumes \cite{Kostyuk}, mean field \cite{Walecka}) as well as specific
conditions of hadronization (e.g., possible conservation of number of quarks
during the hadronization \cite{Zimanyi} ) are rather complicated and model
dependent. So it is naturally to use minimal number of parameters and take
all that into account in the simplest way, i.e., as a unique factor changing
the particle numbers. In other words, we describe the system at the stage of
hadronization as a mixture of ideal gases with additional common chemical
potential $\mu _{ch}$. Afterwards the system expands hydrodynamically with
frozen chemical composition, cools down and finally decays at some thermal
freeze-out hypersurface $\sigma ^{th}(x)$ which could be different for
different particle species \cite{Shuryak,Pratt,Bass}. Then long-lived
particles stream freely into detectors, resonances decay and also contribute
to spectra of observed particles. We reconstruct a state of the system at
the earlier hadronization stage, find energy and particle number densities
and also estimate roughly the effects of interaction at the hadronization
stage using an approach described in the following.

From the analysis of slopes of transverse momentum spectra and of
interferometry radii for different particle species one can restore the
space-time extensions, collective flow parameters and temperature at the
thermal freeze-out stage, when the thermal system decouples and particle
momentum distributions change no longer. The method of joint analysis of the
spectra and correlations was used in Refs. \cite{Sinyukov,Xu,Wiedemann,Ster}
to reconstruct the thermal freeze-out stage. Note that, to evaluate the
detailed behavior of the (pion) spectrum in the whole transverse momentum
region, one needs in detail knowledge of the unobserved resonance yields.
They can be estimated from the temperature and baryochemical potential at
chemical freeze-out. An attempt to determine the resonance yields in that
way was undertaken in Ref. \cite{Wiedemann}. There however, a trajectory
assuming isentropic expansion in full chemical equilibrium was employed
following Ref. \cite{Shuryak}, in contradistinction to the assumption of
chemical freeze-out. In this paper we make a new analysis of the thermal
freeze-out stage within the hydrodynamic approach for lead-lead collisions
based on recent experimental data on one particle spectra and HBT
correlations for {\it different } hadron species. We estimate the basic
parameters at thermal freeze-out from an analysis of the slopes of particle
spectra and of the correlation functions at regions of high enough
transverse momentum to minimize the influence of resonance decays. We use
also the previous results from Refs.\cite{Wiedemann,Ster} to determine the
influence of uncertainties in reconstruction of the thermal freeze-out stage
on the reconstruction of the hadronic gas parameters at chemical freeze-out.

Recent evidence \cite{P. Br.-M.-1,Stachel,P. Br.-M.-2,P. Br.-M.-3} support
strongly the early interpretation \cite{P. Br.-M.-4} that chemical
freeze-out at SPS and higher energies takes place at or near to the phase
boundary (see also the discussion in \cite{Stock}). We employ a similar
approach here with emphasis on the analysis of particle ratios near
mid-rapidity. To complete the reconstruction of the hadronization stage and
find its space-time extension, we solve the hydrodynamic equations with the
Cauchy conditions that are found at thermal freeze-out, and consider the
solution backwards in time until the chemical freeze-out temperature is
reached. The method leads to unique results due to the scale invariance of
the hydrodynamic equations for a mixture of ideal Boltzmann gases with
reference to scale-transformations of pressure and densities of energy and
particles at a fixed temperature. For such a system the variation of initial
densities by a unique factor results in the same scale transformation of the
solutions. Therefore, the temperature as function of time does not depend on
the common chemical potential, which defines absolute values of densities:
the temperature depends only on concentrations of different particles
species. The latter do not change during the hydrodynamic evolution and are
determined by the baryonic and strangeness chemical potentials and
temperature just after hadronization. Finally, from reconstructed space-time
extension of the chemical freeze-out and a particle numbers in mid-rapidity
we determine the densities at hadronization and the common chemical
potential.

An alternative way to determine hadron gas parameters is used in Ref. \cite
{Shuryak}. There, a hydrodynamic approach is followed from the initial
condition of an equilibrated QGP through the phase transition and
hadronization. Subsequent to hadronization the system is assumed, in fact,
to evolve as a mixture of mutually non-interacting gases with fixed chemical
composition. In the present approach we, instead, solve the hydrodynamic
equations for a complete system under the explicit condition of particle
numbers conservation. The distinctive feature of our method is the analysis
backwards in time.

One more important point which we would like to comment here concerns the
question of applicability of the hydrodynamic approach to post hadronization
stage in A+A collisions. In recent papers \cite{Bass,Teaney} it is stated
that the hydrodynamic approach is unable to describe the stage of hadron
phase expansion in heavy ion collisions. Note, however, that the criticism
of Refs. \cite{Bass,Teaney} as to the hydrodynamics is based on calculations
of chemically equilibrium evolution of the mixture of hadron gases. This
evolution differs essentially from the evolution of chemically frozen gas
\cite{Hirano,Teaney1}. In very recent work \cite{Teaney1} the first detailed
comparisons of RQMD calculations with chemically frozen hydrodynamics were
done. There it was demonstrated that if chemical freeze-out is incorporated
into the hydrodynamics, then the final spectra and fireball lifetimes are
insensitive to the temperature at which the switch from hydrodynamics to
cascade is made. It implies that the transformation of heat energy into
collective flow for a chemically frozen hydrodynamic evolution results in
spectra which correspond approximately to the microscopic cascade
calculations. Moreover, the initial conditions of hydro-expansion in Refs.
\cite{Bass,Teaney} correspond to chemically equilibrium mixture of Boltzmann
hadronic gases and does not include possible specific effects of
hadronization process, excluded volumes and collective interaction. The
estimation of these effects, which may reducing the particle number, is one
of the important aims of this paper. Therefore, the criticism in Refs \cite
{Bass,Teaney} as to applicability of hydrodynamics is based on the
hydrodynamic models which differ essentially from what is used in our paper.
The opposite to Refs. \cite{Bass,Teaney} conclusion as to applicability of
hydrodynamics to the post hadronization stage in A+A collisions, was argued
recently in Refs. \cite{Kolb}. There it was shown, in particular, that the
hydrodynamics describes good the single particle spectra and elliptic flows
also in RHIC experiments.

Nevertheless, the cascade models \cite{Pratt,Bass,Teaney} of post
hadronization stage are considered now as favorable alternative to the
hydrodynamic approach. The picture of evolution and particle emission are
rather different in hydrodynamic approach and cascade models (RQMD, UrQMD,
etc...). While in standard hydrodynamic picture the particle spectra are
evaluated usually according to the Cooper-Frye prescription \cite{Cooper}
(using the freeze-out hypersurfaces), the cascade calculations does not
demonstrate the sharp region of particle emission \cite{Pratt}. It does not
mean, however, that hydrodynamic description is incorrect: the absence of
principal contradictions between the both methods of spectra calculations
was demonstrated quite recently in Ref. \cite{Akkelin} based on exact
solution of Boltzmann equation. The method of escape probabilities, proposed
there, gives possibility to make bridge between hydrodynamic and transport
models for rarefied systems. As to rather dense systems, similar to those
that form just after hadronization, where the time of collision is similar
to the time between collisions and collective effects, such as multiparticle
collisions and modification of hadron properties, may take place, the
oversimplified version of transport approach, like UrQMD, cannot be applied
correctly while hydrodynamic approach is more suitable. Note also that at
CERN SPS and RHIC energies the cascade models fail actually to describe the
interferometry radii of the system (see, e.g., \cite{Ferenc,Teaney}).

In Sec. II we discuss the possibility to extract the thermodynamic
parameters at chemical freeze-out from the particle number ratios when those
parameters are non-uniformly distributed in rapidity. In Sec. III we
consider the particle number ratios in a hydrodynamic approach to A+A
collisions. The optimization procedure is applied to fit recent experimental
data for Pb+Pb (CERN, SPS) collisions and find the temperature, baryonic and
strangeness chemical potentials at chemical freeze-out. In Sec. IV we
analyze the property of hydrodynamic expansion in a central slice of
space-time rapidity for a chemically frozen mixture of ideal Boltzmann
gases. Approximate hydrodynamic solutions for non-relativistic transverse
flows are found. In Sec.V the Cauchy conditions at thermal freeze-out, such
as proper time, transverse Gaussian radius parameter, temperature and
transverse velocities are defined from the analysis of spectra and HBT
correlations. In Sec.VI the corresponding `` inverse in time'' hydrodynamic
solution is analyzed and the space-time characteristics of the matter at the
hadronization stage are reconstructed. The later are applied to restore the
energy and particle number densities at that stage. The conclusions and a
discussion of the results are presented in Sec.VII.

\section{The particle numbers at mid-rapidity for expanding HG}

We assume that just after hadronization, the system can be described at some
hypersurface $\sigma ^{ch}(x)$ as a locally and chemically equilibrated
expanding hadron resonance gas. The effects of interaction and possible
peculiarities of the hadronizaton process are taken into account by a common
factor changing particle numbers in the ideal gas; the factor is associated
with an additional common chemical potential $\mu _{ch}$. In concrete
evaluations, as was performed by many authors, instead of considering a
finite chemical and thermal freeze-out hypersurfaces $\sigma (x)$ in
Minkovsky space, we use the common profile factor $\rho (x)$ at proper time $%
\tau (r)\approx const$ to guarantee the finiteness of the system (see the
discussion on the validity of such an approximation in \cite{Xu}). Then the
total multiplicity of a particle species $i$ at chemical freeze-out is
calculated by

\begin{equation}
N_{i}=\int \frac{d^{3}p}{p^{0}}d\sigma _{\mu }^{ch}(x)p^{\mu }f\left( \frac{%
p_{\mu }u^{\mu }(x)}{T_{ch}(x)},\frac{\mu _{i,ch}(x)}{T_{ch}(x)}\right) \rho
(x),  \label{N-tot-1}
\end{equation}
where $f(p_{\mu }u^{\mu }(x)/T_{ch}(x),\mu _{i,ch}(x)/T_{ch}(x))$ is the
Bose-Einstein or Fermi-Dirac distribution function in our approach , and $%
\mu _{i,ch}=B_{i}\mu _{B}+S_{i}\mu _{S}+\mu _{ch}$. The additional chemical
potential $\mu _{ch}$ is unique for all particles species. The effect of
isotopic chemical potential has been investigated in \cite{P. Br.-M.-2} and
found to be small. It is neglected in this paper. Taking into account the
invariance of the distribution function under Lorentz transformation, one
can carry out integrations over momentum variables and finally get
\begin{equation}
N_{i}=\int d\sigma _{\mu }^{ch}(x)u^{\mu }(x)n_{i}^{ch}(x),\quad
n_{i}^{ch}(x)=\overline{n}_{i}^{ch}(x)\rho (x)  \label{N-total}
\end{equation}
where in Boltzmann approximation for the distribution function the
thermodynamic densities $\overline{n}_{i}^{ch}(x)$ have the form
\begin{equation}
\overline{n}_{i}^{ch}(x)=\frac{(2J_{i}+1)}{2\pi ^{2}}T_{ch}(x)m_{i}^{2}\exp
(\mu _{i,ch}(x)/T_{ch}(x))K_{2}(m_{i}/T_{ch}(x)),  \label{eq:1}
\end{equation}
and $K_{n}(u)=\frac{1}{2}\int_{-\infty }^{+\infty }dz\exp [-u\cosh z+nz]$ ,
with $%
%TCIMACRO{\func{Re}}%
%BeginExpansion
\mathop{\rm Re}%
%EndExpansion
u>0,$ is the modified Bessel function of order $n$ ($n=0,1,...$). One can
rewrite (\ref{N-total}) in the form
\begin{equation}
N_{i}=\langle \overline{n}_{i}^{ch}\rangle V^{ch}  \label{N-V}
\end{equation}
where the volume $V^{ch}$ of system is expressed through the integral of the
hydrodynamic $4$ -velocity $u^{\mu }(x)$ and the profile factor $\rho (x)$
over the hypersurface $\sigma ^{ch}$:
\begin{equation}
V^{ch}=\int d\sigma _{\mu }^{ch}(x)u^{\mu }(x)\rho (x).  \label{v}
\end{equation}
We further define the density averaged over the chemical freeze-out
hypersurface by
\begin{equation}
\langle \overline{n}_{i}^{ch}\rangle =\frac{\int d\sigma _{\mu }^{ch}u^{\mu
}(x)\overline{n}_{i}^{ch}(x)\rho (x)}{\int d\sigma _{\mu }^{ch}u^{\mu
}(x)\rho (x)}.  \label{eq:02}
\end{equation}
If temperature and chemical potentials are uniform over the chemical
freeze-out hypersurface, then $\langle \overline{n}_{i}^{ch}\rangle =%
\overline{n}_{i}^{ch}(T_{ch},\mu _{i,ch})$ and the standard analysis of
particle number ratios in $4\pi $ geometry can be performed to find the
temperature and chemical potentials. However, if $T$ and $\mu $ are not
uniformly distributed over the hydrodynamic tube at\ the hypersurface $%
\sigma ^{ch}$, then the similar analysis implying the substitution $\langle
\overline{n}_{i}\rangle \rightarrow \overline{n}_{i}^{ch}(\langle
T_{ch}\rangle ,\langle \mu _{i,ch}\rangle )$ with some average temperature
and chemical potentials, could result in significant deviations for the
values one is interested in \cite{Sollfrank}. To reduce the corresponding
uncertainties, the analysis of particle ratios in a narrow rapidity interval
near mid-rapidity should be used as we will discuss below.

To analyze non-uniformly distributed systems, let us start from the well
known observation \cite{Landau} that, at the latest stage of a hydrodynamic
evolution, the longitudinal velocity distribution corresponds to the
asymptotic quasi-inertial regime, $v_{L}=z/t$. This property, which is the
ansatz in the Bjorken model \cite{Bjorken} for each moment of time, is
realized at the freeze-out stage even in the Landau model of complete
stopping \cite{Landau}.\footnote{%
The asymptotic quasi-inertial property one can see also in the analytic
solutions for elliptic flow in non-relativistic approximation \cite{Csorgo}.}
Consequently, we assume in the following the boost-invariance for
longitudinal hydrodynamic velocities at the last (hadronic) stage of the
evolution, $v_{L}=z/t$. This then allows us to parameterize $t$ and $z$ at $%
\sigma $ by the form: $t=\tau (r,\eta )\cosh (\eta )$, $z=\tau (r,\eta
)\sinh (\eta )$, where $r$ is absolute value of the transverse coordinate $%
{\bf {r}}$ and $\eta $ is the longitudinal fluid rapidity, $\eta =\tanh
^{-1}v_{L}$. The parameter $\tau (r,\eta )$ generalizes the Bjorken proper
time parameter $\tau $. Taking into account also a transverse velocity
component ($v(r,\tau )$ in the longitudinally co-moving system, LCMS), we
obtain for the 4-velocity

\begin{equation}
u^{\mu }(r,\eta )=\gamma \left( \cosh \eta ,v\cos \phi ,v\sin \phi ,\sinh
\eta \right) ,  \label{transv}
\end{equation}
where $\gamma =(1-v^{2})^{-1/2}$. The element of the hypersurface $\sigma
(x) $ takes the form

\begin{equation}
d\sigma _{\mu }=\tau (r,\eta )d\eta dr_{x}dr_{y}\left( \frac{1}{\tau }\frac{%
d\tau }{d\eta }\sinh \eta +\cosh \eta ,-\frac{d\tau }{d{r}_{x}},-\frac{d\tau
}{d{r}_{y}},-\frac{1}{\tau }\frac{d\tau }{d\eta }\cosh \eta -\sinh \eta
\right) .  \label{sigm}
\end{equation}

Let us, first, introduce for an expanding HG the effective volume $V_{eff}$
attributed to one unit of rapidity. To determine $V_{eff}$ one can write the
particle number distribution differential in fluid rapidity $\eta $ for
particle of species $"i"$ at chemical freeze-out:
\begin{equation}
\frac{dN_{i}}{d\eta }=\int \frac{d^{3}p}{p^{0}}\frac{d\sigma _{\mu }^{ch}}{%
d\eta }p^{\mu }f\left( \frac{p_{\mu }u^{\mu }(r,\eta )}{T_{ch}(\eta )},\frac{%
\mu _{i,ch}(\eta )}{T_{ch}(\eta )}\right) \rho (r,\eta ).  \label{dn-deta}
\end{equation}
Here we assume that the chemical potentials as well as the temperature are
constants in transverse direction across the freeze-out hypersurface. Taking
into account the invariance of the distribution function under Lorentz
transformation, one can carry out integrations over momentum variables and
finally gets
\begin{equation}
\frac{dN_{i}}{d\eta }=V_{eff}^{ch}(\eta )\overline{n}_{i}^{ch}(\eta )
\label{eq:0.1}
\end{equation}
and
\begin{equation}
V_{eff}^{ch}(\eta )=\int \frac{d\sigma _{\mu }^{ch}}{d\eta }u^{\mu }(r,\eta
)\rho (r,\eta ).  \label{eq:0.2}
\end{equation}

The particle number distribution over momentum rapidity $y$ is
\begin{equation}
\frac{dN_{i}}{dy}=\int \frac{d^{3}p}{p^{0}dy}d\sigma _{\mu }^{ch}p^{\mu
}f\left( \frac{p_{\mu }u^{\mu }(r,\eta )}{T_{ch}(\eta )},\frac{\mu
_{i,ch}(\eta )}{T_{ch}(\eta )}\right) \rho (r,\eta ).  \label{eq:1.1}
\end{equation}
One can rewrite Eq. (\ref{eq:1.1}) in the form
\begin{equation}
\frac{dN_{i}}{dy}=\frac{dN_{i}}{d\eta }\mid _{\eta =y}(1+\delta
_{i}(y))=V_{eff}^{ch}(y)(1+\delta _{i}(y))\overline{n}_{i}^{ch}(y)
\label{eq:03}
\end{equation}
where
\begin{equation}
\delta _{i}(y)=\frac{\frac{dN_{i}}{dy}-\frac{dN_{i}}{d\eta }\mid _{\eta =y}}{%
\frac{dN_{i}}{d\eta }\mid _{\eta =y}}  \label{delta}
\end{equation}
From Eq. (\ref{eq:03}), one can conclude that, for a nonuniform distribution
of thermodynamic parameters along the beam axis, the analysis of particle
number ratios at mid-rapidity, $y\approx 0$ can be based on ratios $%
\overline{n}_{i}/\overline{n}_{j}$ if correction factors $\delta _{i}(0)$
are small. One can include corresponding corrections in the non-unique
``effective volume'' $V_{eff,i}^{ch}=V_{eff}^{ch}(1+\delta _{i})$ which is
different for different particle species.

Let us estimate the correction factors. The saddle-point approximation to
the integrals over fluid rapidity $\eta $ in (\ref{eq:1.1}) can be applied
when the parameter $m_{i}/T(\eta )\gg 1$ at the saddle point $\eta \approx y$%
. Then, at mid-rapidity, the area around $\eta =y=0$ gives the main
contribution to the integral over $\eta $ in (\ref{eq:1.1}) and one can
expand the slowly varying functions into a Tailor series around the
saddle-point $\eta =0$.{\bf \ }For the sake of simplicity, we neglect a
variation of $T(\eta )$ in space rapidity $\eta $ in comparison with the
variation of $\mu _{i}^{ch}(\eta )$, assuming that all non-uniformity
originated from incomplete baryonic stopping. This implies an increase of
the baryonic chemical potential with the absolute value of rapidity.
Assuming that the transverse expansion is non-relativistic, with a
freeze-out that occurs at approximately equal proper time $\tau $, and that
the profile $\rho (x)$ can be factorized, $\rho (x)=\rho _{T}(r)\rho
_{L}(\eta )$, we estimate the correction factor at mid-rapidity:

\begin{equation}
\delta _{i}\approx \frac{1}{2}\frac{(\rho _{L}\exp (\mu _{i,ch}/T_{ch})\tau
_{ch})^{\prime \prime }\mid _{\eta =0}}{\rho _{L}(0)\exp (\mu
_{i,ch}/T_{ch})\tau _{ch}}\frac{T_{ch}}{m_{i}}  \label{correction}
\end{equation}
All thermodynamic parameters here are taken at mid-rapidity. One can see
that the factor $\delta _{i}$ is rather small for baryons, because the ratio
$T_{ch}/m_{i}$ is small, as well as it is small for pions, because their
baryonic number are equal to zero. Therefore, as follows from Eq.(\ref{eq:03}%
) for $\delta _{i}\ll 1$ the analysis of particle number ratios at a narrow
mid-rapidity interval can be based on ratios $\overline{n}_{i}^{ch}/%
\overline{n}_{j}^{ch}$ \cite{Sinyukov}. These values are the same as for
Boltzmann gas in a static volume \cite{Sinyukov,Cleymans}.

It is worth to note that the second derivative in formula (\ref{correction})
is positive for incomplete baryonic stopping and, so, $\delta _{i}>0$. As we
will discuss later, the consequences of it are non-trivial and rather
important for pion production via decay of baryonic resonances after the
thermal freeze-out occurs.

\section{The analysis of particle number ratios}

In ultrarelativistic A+A collisions a hadron resonance gas is assumed to
appear due to the hadronization process. Subsequent, it expands and reaches
the final stage of the evolution - the thermal freeze-out. According to the
hypotheses of chemical freeze-out, the total number of each particle species
is preserved during the evolution. After the hydrodynamic tube decays and
produces final particles, one has to take into account also particles that
come from resonance decays. In our calculations the resonance mass spectrum
extends over all mesons and baryons with masses below $2$ GeV. Decay
cascades are also included in the analysis. The value of the strangeness
chemical potential $\mu _{S}$ is fixed by baryonic chemical potential and
temperature from the condition of local strangeness conservation. We
optimize the particle number ratios at mid-rapidity, using the corresponding
values for Pb + Pb collisions at 158A GeV/c from WA97 (Table 1) and NA49
(Table 2) experiments, excluding from optimization procedure pions as the
lightest particles, the later circumstance can result in serious
corrections, besides of Bose-Einstein enhancement effect in the distribution
function, which we will discuss below. Then we describe the data in the
model of a chemically equilibrated ideal hadron gas (HG). The feeding in our
calculations is tuned according to the experimental conditions as described
below.

In the geometry of the WA97 experiment the feed-down from weak cascade
decays is expected to be of minor importance \cite{WA97}, and so the $%
\Lambda $ ($\overline{\Lambda }$) have not been corrected for feed-down from
$\Xi $ weak decays, and $\Xi $ have not been corrected for feed-down from $%
\Omega $. But feeding of $\Lambda $ ($\overline{\Lambda }$) from
electromagnetic decays of $\Sigma ^{0}$ ($\overline{\Sigma }^{0}$) is
included. We did not include (according to the NA49 experimental procedure
\cite{NA49-4}) feeding of $p$ and $\overline{p}$ from $\overline{\Lambda },$
$\Lambda $ and from $\Sigma ^{+},\overline{\Sigma }^{+}$. In the NA49
experiment $\Lambda $ and $\overline{\Lambda }$ yields include the
contributions of $\Sigma ^{0}$ and $\overline{\Sigma }^{0}$ as well as
feed-down from weak $\Xi $ decays \cite{NA49-4}. The feeding of $\pi ^{-}$
from kaons and from $\Lambda $ is not included \cite{NA49-1}. We included
feeding from $\phi $ decay into kaons (NA49 and WA97) and pions (NA49). For
technical reasons in the model calculations we used the following
substitution: $K_{s}^{0}/\pi ^{-}\rightarrow (K^{-}+K^{+})/2\pi ^{-}$, a
good approximation for Pb+Pb collisions \cite{NA49-2}. The results for the
best fit are shown in Tables 1 and 2 for Boltzmann statistic (BS) and
quantum statistic (QS) calculations. The obtained temperatures $%
T_{ch}\approx 164$ MeV and baryochemical potential $\mu _{B}\approx 224$ MeV
differ slightly from the corresponding values of Ref. \cite{P. Br.-M.-2}, $%
T_{ch}\approx 168$ MeV, $\mu _{B}\approx 266$ MeV, that were gotten from
analysis of particle number ratios in different limited rapidity intervals
as well as in full $4\pi $-geometry.

TABLE 1. Experimental particle number ratios (WA97) in Pb+Pb SPS collisions
, compared to model calculation for Boltzmann statistic (BS) and quantum
statistic (QS)

\begin{center}
\begin{tabular}{|l|l|l|l|}
\hline
$T_{ch}$ &  & $163.37$ MeV & $163.91$ MeV \\ \hline
$\mu _{B}$ &  & $223.6$ MeV & $224.0$ MeV \\ \hline
$\mu _{S}$ &  & $55.2$ MeV & $55.1$ MeV \\ \hline
Ratios & Data & BS & QS \\ \hline
$\overline{\Lambda }/\Lambda $ & $0.133\pm 0.007$ \cite{WA97} & $0.133$ & $%
0.133$ \\ \hline
$\Xi ^{+}/\Xi ^{-}$ & $0.249\pm 0.019$ \cite{WA97} & $0.250$ & $0.249$ \\
\hline
$\Xi ^{-}/\Lambda $ & $0.100\pm 0.004$ \cite{WA97} & $0.098$ & $0.099$ \\
\hline
$\Xi ^{+}/\overline{\Lambda }$ & $0.188\pm 0.016$ \cite{WA97} & $0.186$ & $%
0.186$ \\ \hline
$\Omega ^{-}/\Xi ^{-}$ & $0.182\pm 0.022$ \cite{WA97} & $0.139$ & $0.140$ \\
\hline
$\Omega ^{+}/\Xi ^{+}$ & $0.281\pm 0.053$ \cite{WA97} & $0.274$ & $0.274$ \\
\hline
$\Omega ^{+}/\Omega ^{-}$ & $0.383\pm 0.081$ \cite{WA97} & $0.492$ & $0.489$
\\ \hline
$\Lambda /K_{s}^{0}$ & $0.6706$ \cite{Evans} & $0.6212$ & $0.6226$ \\ \hline
\end{tabular}
\end{center}

TABLE 2. Experimental particle number ratios (NA49) in Pb+Pb SPS collisions
, compared to model calculation for Boltzmann statistic (BS) and quantum
statistic (QS).

\begin{center}
\begin{tabular}{|l|l|l|l|}
\hline
$T_{ch}$ &  & $163.37$ MeV & $163.91$ MeV \\ \hline
$\mu _{B}$ &  & $223.6$ MeV & $224.0$ MeV \\ \hline
$\mu _{S}$ &  & $55.2$ MeV & $55.1$ MeV \\ \hline
Ratios & Data & BS & QS \\ \hline
$(p-\overline{p})/\pi ^{-}$ & $0.184$ \cite{NA49-1} & $0.245$ & $0.239$ \\
\hline
$(p-\overline{p})/K^{-}$ & $1.776$ \cite{NA49-1,NA49-2} & $1.847$ & $1.843$
\\ \hline
$(p-\overline{p})/K^{+}$ & $1.028$ \cite{NA49-1,NA49-2} & $1.090$ & $1.086$
\\ \hline
$K^{+}/K^{-}$ & $1.727$ \cite{NA49-2} & $1.695$ & $1.697$ \\ \hline
$\overline{p}/p$ & $0.115$ \cite{NA49-1} & $0.066$ & $0.067$ \\ \hline
$\phi /(p-\overline{p})$ & $0.0802$ \cite{NA49-1,NA49-2,Sikler} & $0.0937$ &
$0.0935$ \\ \hline
\end{tabular}
\end{center}

The deviation between calculated particle ratios and experimental data are
within the error of the data of WA97, excepts for the ratios which contain $%
\Omega $. As to the $\overline{p}/p$ ratio of NA49, one has take into
account that the corresponding data $\overline{p}/p\approx 0.11$ are rather
preliminary and differ much from published NA44 data where the value is
twice less \cite{NA44}. It is interesting to note, that in rapidity interval
$2.0<y<2.5$ shifted by 0.65 units of rapidity from central mid-rapidity
point, the correspondent value measured by NA49 \cite{NA49-6} is $0.06\pm
0.01$.

As one can see from Table 2 the most significant deviation between model and
data is for the $(p-\overline{p})/\pi ^{-}$ ratio where the calculations
predict\ a value about 30 percent larger then observed. Considering typical
sizes of systematic errors in the data of at least 10 percent one should,
however, not overestimate the statistical relevance of the discrepancy. In
part, it maybe due to the reduction, compared to the results of Ref.\cite{P.
Br.-M.-2}, in $\mu _{B}$ since there are less baryons at mid-rapidity then
are expected in scenario of full baryon stopping. Because about 1/2 of the
pions result from baryonic resonance decays this also leads to a reduction
of pions in the model and, thereby, over-prediction of the $(p-\overline{p}%
)/\pi ^{-}$ ratio.

The above effects of the non-uniformity in rapidity of the baryochemical
potential can be estimated numerically. First, note that our result for $(p-%
\overline{p})/\pi ^{-}$ ratio implies the same effective volume for pions as
for other heavier hadrons. In fact, the pion volume is larger due to pion
contribution from resonance decays. To explain that, let us, first, note
that because of large resonance masses, comparing with the thermal
freeze-out temperature, the momentum rapidity $y$ of a resonance coincides
with the rapidity $\eta $ of the corresponding decaying fluid element and,
therefore, the distribution of resonances in $y$ follows their distribution
in space-time rapidity $\eta $. The latter is determined mainly by the
baryochemical potential that has a shallow minimum at $\eta =0$. When
resonances decay into the lightest particles, pions, there is a rather wide
kinematic ``window'' $\Delta y$ in rapidity for emitted pions. Then
resonances with $y$ beyond mid-rapidity may affect the multiplicities of
pions at mid-rapidity. It is similar to a diffusion process: there is
transport of pions from baryon rich peripheral $y$- regions to the central
region. One can roughly estimate the effect of pion ``diffusion'' towards
mid- rapidity using an effective temperature $T_{dec}$ for those pions
resulting from baryon decays. Taking into account that $m_{\pi }/T_{dec}\sim
1$, one can see from Eqs.(\ref{eq:03}) and (\ref{correction}) with
substitutions $\rho _{L}\exp (\mu _{i,ch}/T_{ch})\tau _{ch}\rightarrow \rho
_{L}^{dec}\tau _{th}$, $T_{ch}/m_{i}\rightarrow T_{dec}/m_{\pi }$, that a
correction, increasing the effective volume for pions produced by baryonic
resonances, could appear. Our numerical estimate is based on the baryonic
density profile for AGS energies taken from Ref.\cite{Shengqin}, gives a
contribution of about $10$ $\%$ for pion transport to mid-rapidity region.
Whether there are other mechanisms increasing, in particular, the number of
pions at mid-rapidity, such as decay of a potentially formed disoriented
chiral condensate, cannot be decided by the present analysis.

To provide the hydrodynamic calculations and estimate the energy density one
may, phenomenologically, ``fix'' $(p-\overline{p})/\pi ^{-}$ ratio by
introducing an additional chemical potential for direct pions at chemical
freeze-out of $\mu _{\pi }=89$ MeV. Note, however, that the apparent anomaly
in $(p-\overline{p})/\pi ^{-}$ ratio has only little influence on the
hydrodynamic behavior of chemically frozen HG considered in the next section.

\section{Hydrodynamic Evolution of Chemically Frozen HG}

We describe hadron matter evolution at mid-rapidity after hadronization by
the relativistic hydrodynamics equations for an ideal fluid:
\begin{eqnarray}
\partial _{\mu }T^{\mu \nu } &=&0,  \label{eq:3.1} \\
\partial _{\mu }u^{\mu }n_{i} &=&0,  \label{eq:3.2}
\end{eqnarray}
where $T^{\mu \nu }=(\varepsilon +p)u^{\mu }u^{\nu }-g^{\mu \nu }p$ is the
energy-momentum tensor of the Boltzmann-like gas (see Sec. II), $\varepsilon
=\sum \varepsilon _{i}$ is energy density and $p=\sum p_{i}$ is pressure.
The energy densities are proportional to the profile factor $\rho (x)$
similar to density of particle species $"i"$ (\ref{N-total}): $\varepsilon
_{i}(x)=\rho (x)\overline{\varepsilon }_{i}(x)$. Here $\overline{\varepsilon
}_{i}(x)$ are thermodynamic energy densities in the Boltzmann gas. The
equation of state for such a system has the form:
\begin{equation}
p=nT,\quad n=\sum n_{i}.  \label{eq:4}
\end{equation}
Due to chemical freeze-out the conservation of particle number density
currents for each species of hadrons is explicitly implemented in Eqs. (\ref
{eq:3.2}).\ We ignore there the influence of possible changes of particle
numbers due to decay of resonances to the time dependence of the macroscopic
(hydrodynamic) variables, the validity of such an approximation will be
discussed below in Sec. VI.

We assume a quasi-inertial, Bjorken-like, longitudinal flow with velocity $%
z/t$ at the post-hadronization stage of evolution. Thereby 4-velocities are
described by Eq. (\ref{transv}). Using this flow ansatz and assuming that
the hypersurfaces $\sigma (x)$, where the flow is in a quasi-inertial regime
(\ref{transv}), correspond to the proper time $\tau (r,\eta )=\sqrt{%
t^{2}-z^{2}}\approx const$, one can rewrite energy-momentum conservation
equations (\ref{eq:3.1}) in the form \cite{Biro}:
\begin{eqnarray}
\left( \frac{\partial }{\partial \tau }+v\frac{\partial }{\partial r}\right)
\varepsilon +(\varepsilon +p)D &=&0,  \label{eq:3.3} \\
\gamma ^{2}(\varepsilon +p)\left( \frac{\partial }{\partial \tau }+v\frac{%
\partial }{\partial r}\right) v+\left( \frac{\partial }{\partial r}+v\frac{%
\partial }{\partial \tau }\right) p &=&0,  \label{eq:3.4} \\
\frac{1}{\tau }\frac{\partial }{\partial \eta }p &=&0,  \label{eq:3.5} \\
\frac{1}{r}\frac{\partial }{\partial \phi }p &=&0,  \label{eq:3.6}
\end{eqnarray}
where
\begin{equation}
D=\gamma ^{2}\left( \frac{\partial }{\partial r}+v\frac{\partial }{\partial
\tau }\right) v+\frac{1}{\tau }+\frac{v}{r}.  \label{eq:3.7}
\end{equation}
In a similar way particle number conservation equations (\ref{eq:3.2}) can
be rewritten in the form
\begin{equation}
n_{i}\left( \frac{\partial \gamma }{\partial \tau }+\frac{\partial }{%
\partial r}(\gamma v)+\gamma \left( \frac{1}{\tau }+\frac{v}{r}\right)
\right) +\gamma \left( \frac{\partial }{\partial \tau }+v\frac{\partial }{%
\partial r}\right) n_{i}=0.  \label{eq:3.8}
\end{equation}

Note that the equations (\ref{eq:3.3}) - (\ref{eq:3.8}) are related, in
general, to non-uniformly distributed energy and particle densities. Among
them, the equation (\ref{eq:3.5}) is the only one that governs the
dependence of densities on fluid rapidity $\eta $. The others contain $\eta $
as a parameter. It allows one to analyze the solutions of the hydrodynamic
equations (\ref{eq:3.3}) - (\ref{eq:3.8}) for any fixed rapidity. Our
special interest concerns the point $\eta \approx 0$ corresponding to the
main contribution to multiplicities at central rapidity. As we are
interested primarily in the bulk properties of the system, we will search
for the solutions of Eqs. (\ref{eq:3.3}) - (\ref{eq:3.8}) in the center of
system, using a following approximations for transverse flow:
\begin{eqnarray}
v^{2} &\ll &1,  \label{nonrelat1} \\
\quad \left| v\frac{\partial v}{\partial \tau }\right| &\ll &\frac{\partial v%
}{\partial r}.  \label{nonrelat2}
\end{eqnarray}
The condition (\ref{nonrelat1}) is just a non-relativistic approximation for
the transverse velocity (in LCMS for fluid element it coincides with $v$),
whereas Eq. (\ref{nonrelat2}), as it will be discussed below, can be
justified by a relatively large average energy per particle in a hadron
resonance gas. Using the conditions (\ref{nonrelat1}) and (\ref{nonrelat2}),
one can get from Eqs. (\ref{eq:3.3}),\ (\ref{eq:3.4}) and (\ref{eq:3.8}) the
approximate hydrodynamic equations
\begin{eqnarray}
\left( \frac{\partial }{\partial \tau }+v\frac{\partial }{\partial r}\right)
\varepsilon +(\varepsilon +p)\left( \frac{\partial v}{\partial r}+\frac{1}{%
\tau }+\frac{v}{r}\right) &=&0,  \label{eq:5} \\
(\varepsilon +p)\left( \frac{\partial }{\partial \tau }+v\frac{\partial }{%
\partial r}\right) v+\left( \frac{\partial }{\partial r}+v\frac{\partial }{%
\partial \tau }\right) p &=&0,  \label{eq:5.01} \\
n_{i}\left( \frac{\partial v}{\partial r}+\frac{1}{\tau }+\frac{v}{r}\right)
+\left( \frac{\partial }{\partial \tau }+v\frac{\partial }{\partial r}%
\right) n_{i} &=&0.  \label{eq:5.02}
\end{eqnarray}

Let us, motivated by the initial conditions at hadronization stage as well
as exact analytical solution of non-relativistic hydrodynamics \cite{Csorgo}
and Bjorken boost-invariant solution \cite{Bjorken} (see also \cite{Grassi}%
), assume the following ansatz for\ densities $n_{i}$, transverse velocity $%
v $ and temperature $T$ in a central slice of space-time rapidity $\eta
\approx 0$:
\begin{eqnarray}
n_{i}(\tau ,r) &=&n_{i}(\tau _{0},0)\frac{R_{T}^{2}(\tau _{0})\tau _{0}}{%
R_{T}^{2}(\tau )\tau }\rho (\tau ,r),  \label{eq:5.03} \\
\rho (\tau ,r) &=&\exp \left( -\frac{r^{2}}{2R_{T}^{2}(\tau )}\right) ,
\label{eq:5.04} \\
v(\tau ,r) &=&\frac{r}{R_{v}(\tau )},  \label{eq:5.05} \\
T(\tau ,r) &=&T(\tau ).  \label{eq:5.06}
\end{eqnarray}
Here and below we omit indication of $\eta $ - dependence of hydrodynamic
variables. Using the above ansatz for the densities (\ref{eq:5.03}), (\ref
{eq:5.04}) and flow velocity distribution (\ref{eq:5.05}), we find that the
particle number conservation equation (\ref{eq:5.02}) is satisfied if
\begin{equation}
\frac{1}{R_{v}}=\frac{\stackrel{.}{R}_{T}}{R_{T}}.  \label{eq:5.07}
\end{equation}
Here and hereafter we designate by dotted symbols the derivatives in proper
time: $\stackrel{.}{f}=df/d\tau $.

Using the equation of state (\ref{eq:4}) and conservation of particle number
density currents (\ref{eq:5.02}), one can rewrite Eq. (\ref{eq:5}) in the
form:
\begin{equation}
\frac{\partial e(T)}{\partial T}\frac{\partial T}{\partial \tau }+T\left(
\frac{\partial v}{\partial r}+\frac{1}{\tau }+\frac{v}{r}\right) =0,
\label{eq:5.21}
\end{equation}
where

\begin{equation}
\quad e(T)=\sum \kappa _{i}e_{i}(T),\quad \kappa _{i}\equiv \frac{n_{i}}{n}=%
\frac{\overline{n}_{i}^{ch}}{\overline{n}^{ch}},\quad e_{i}(T)\equiv \frac{%
\varepsilon _{i}}{n_{i}}=m_{i}\frac{K_{1}(m_{i}/T)}{K_{2}(m_{i}/T)}+3T.
\label{CV-def}
\end{equation}
Using the ansatz for transverse velocity distribution and temperature (\ref
{eq:5.05}) - (\ref{eq:5.07}), we get from Eq. (\ref{eq:5.21}) a first order
ordinary differential equation for the transverse Gaussian radius parameter $%
R_{T}$ and temperature $T$:
\begin{equation}
\frac{\partial e(T)}{\partial T}\stackrel{.}{T}+T\left( 2\frac{\stackrel{.}{R%
}_{T}}{R_{T}}+\frac{1}{\tau }\right) =0.  \label{eq:8}
\end{equation}
Just for illustration one can solve Eq. (\ref{eq:8}) in the case of {\it %
non-relativistic} ideal Boltzmann gas when $m_{i}/T\gg 1$. Then the energy
per particle is equal to
\begin{equation}
e_{i}(T)\approx m_{i}+3/2T,  \label{eq:8.1}
\end{equation}
and one gets
\begin{equation}
T(\tau )=T(\tau _{0})\left( \frac{R_{T}^{2}(\tau _{0})\tau _{0}}{%
R_{T}^{2}(\tau )\tau }\right) ^{2/3}.  \label{eq:9}
\end{equation}
We will not use such an approximation which is too rough for real
composition of hadron resonance gas.

With the help of the equation of state (\ref{eq:4}) and the ansatz (\ref
{eq:5.05}) - (\ref{eq:5.07}) for transverse velocity distribution and
temperature the equation (\ref{eq:5.01}) is reduced to
\begin{equation}
\left( e(T)+T\right) \stackrel{..}{R}_{T}R_{T}+\stackrel{.}{R}_{T}R_{T}%
\stackrel{.}{T}=T\left( 1+2\stackrel{.}{R}_{T}^{2}+\frac{\stackrel{.}{R}%
_{T}R_{T}}{\tau }-\frac{r^{2}\stackrel{.}{R}_{T}^{2}}{R_{T}^{2}}\right) .
\label{eq:10}
\end{equation}
The last term in the bracket on the right hand side of the above equation is
just $-v^{2} $, as follows from Eqs. (\ref{eq:5.05}) and (\ref{eq:5.07}).
Therefore, using the condition (\ref{nonrelat1}), one can neglect this term
as compared with $1$, and finally get the second order ordinary differential
equation
\begin{equation}
\left( e(T)+T\right) \stackrel{..}{R}_{T}R_{T}+\stackrel{.}{R}_{T}R_{T}%
\stackrel{.}{T}=T\left( 1+2\stackrel{.}{R}_{T}^{2}+\frac{\stackrel{.}{R}%
_{T}R_{T}}{\tau }\right) .  \label{eq:11}
\end{equation}
Together with Eq. (\ref{eq:8}) the above equation represents a system of
ordinary differential equations for the transverse Gaussian radius parameter
$R_{T}$ and temperature $T$. These equations, together with ansatz (\ref
{eq:5.05}) - (\ref{eq:5.07}), correspond to\ the basic hydrodynamic
equations (\ref{eq:3.3}) - (\ref{eq:3.8}) with ideal gas equation of state (%
\ref{eq:4}) and under conditions (\ref{nonrelat1}), (\ref{nonrelat2}).\

In the above mentioned limit of the non-relativistic ideal Boltzmann gas ($%
m_{i}/T\gg 1$) the solution of Eq. (\ref{eq:8}) is given by Eq. (\ref{eq:9})
and so the (proper) time derivative of temperature is
\begin{equation}
\stackrel{.}{T}=-\frac{2}{3}T\left( 2\frac{\stackrel{.}{R}_{T}}{R_{T}}+\frac{%
1}{\tau }\right) .  \label{eq:11.1}
\end{equation}
Then, for the same case of non-relativistic ideal Boltzmann gas, Eq. (\ref
{eq:11}) takes the form
\begin{equation}
\stackrel{..}{R}_{T}R_{T}=\frac{T}{\sum \kappa _{i}m_{i}}\left( 1+\frac{10}{3%
}\stackrel{.}{R}_{T}^{2}+\frac{5}{3}\frac{\stackrel{.}{R}_{T}R_{T}}{\tau }%
\right) .  \label{eq:12}
\end{equation}
If, in addition, one neglects the last two terms on the right hand side of
the above equation , one arrives at
\begin{equation}
\stackrel{..}{R}_{T}R_{T}=\frac{T}{\sum \kappa _{i}m_{i}}.  \label{eq:13}
\end{equation}
As is possible to show, Eqs. (\ref{eq:9}) and (\ref{eq:13}) are {\it exact}
equations for {\it non-relativistic} hydrodynamics with corresponding
initial conditions (and obvious substitution $\tau \rightarrow t$). Here we
just demonstrated the chain of steps which are necessary to pass from our
{\it semi-relativistic }hydrodynamic equations (\ref{eq:8}) and (\ref{eq:11}%
) for given initial conditions to {\it non-relativistic} ones.

Let us now discuss the constraints for our ansatz parameters, that are
forced by the conditions (\ref{nonrelat1}), (\ref{nonrelat2}). So far, as we
will apply our solutions to the center part of system, limited by condition $%
r/R_{T}\lesssim 1$, the first restriction (\ref{nonrelat1}) being applied to
ansatz (\ref{eq:5.05}), (\ref{eq:5.07}) means that
\begin{equation}
\stackrel{.}{R}_{T}^{2}\ll 1.  \label{nonrelat3}
\end{equation}
The second constraint (\ref{nonrelat2}), using \ ansatz (\ref{eq:5.05}), (%
\ref{eq:5.07}) and limitation $r/R_{T}\lesssim 1$, can be rewritten in the
form
\begin{equation}
\left| \stackrel{..}{R}_{T}R_{T}-\stackrel{.}{R}_{T}^{2}\right| \ll 1.
\label{nonrelat4.1}
\end{equation}
and, taking into account (\ref{nonrelat3}), results in
\begin{equation}
\left| \stackrel{..}{R}_{T}R_{T}\right| \ll 1.  \label{nonrelat4.2}
\end{equation}
Finally, expressing $\stackrel{..}{R}_{T}R_{T}$ by the help of Eqs. (\ref
{eq:8}) and (\ref{eq:11}), we get
\begin{equation}
\frac{T}{e(T)+T}\left( 1+2\stackrel{.}{R}_{T}^{2}+\frac{\stackrel{.}{R}%
_{T}R_{T}}{\tau }\right) +\frac{\stackrel{.}{R}_{T}R_{T}}{e(T)+T}T\left( 2%
\frac{\stackrel{.}{R}_{T}}{R_{T}}+\frac{1}{\tau }\right) \left( \frac{%
\partial e(T)}{\partial T}\right) ^{-1}\ll 1.  \label{nonrelat4.3}
\end{equation}
We are interested in applying our solution to the post-hadronization stage
of high energy heavy ion collisions, where typically $\tau \sim R_{T}$.
Therefore, the above constraint is satisfied if the average energy per
particle $e(T)$ is large enough:
\begin{equation}
\frac{T}{e(T)+T}\ll 1.  \label{nonrelat4.4}
\end{equation}
The restrictions (\ref{nonrelat3}) and (\ref{nonrelat4.4}) define the region
of applicability of our semi-relativistic solutions of hydrodynamic
equations (\ref{eq:3.3}) - (\ref{eq:3.8}).

Finally, we reduce the set of partial differential equations of chemically
frozen hydrodynamics (\ref{eq:3.3}) - (\ref{eq:3.8}) with relativistic ideal
gas equation of state (\ref{eq:4}) to the set of ordinary differential
equations for the scale parameter $R_{T}$ and the temperature $T$ ,\ (\ref
{eq:8}) and (\ref{eq:11}), that can be solved by presently available
numerical packages,\ e.g., by the program package Mathematica \cite{Mat}.
The initial conditions for these equations are: Bjorken-like initial
longitudinal flow, where velocity $z/t$ is accompanied by transverse flow,
and Gaussian density profile in transverse direction. Such reduction is
based on approximations (\ref{nonrelat1}), (\ref{nonrelat2}), which lead to
the semi-relativistic hydrodynamic equations (\ref{eq:5}) - (\ref{eq:5.02}),
and, as we will see, are satisfied in central domain $r/R_{T}\lesssim 1$ at
post-hadronization stage of matter evolution in Pb +Pb \ collisions at SPS
CERN.

One can see that the equations (\ref{eq:8}), (\ref{eq:11}) depend only on
the concentrations $\kappa _{i}$ of different particle species and do not
depend on absolute values of densities, and, thus, do not depend on the
unknown common chemical potential $\mu _{ch}$. \ The concentrations $\kappa
_{i}$ are completely defined \ by the temperature and baryochemical
potential that are extracted from the fitting of particle number ratios (see
Sec. III). Then Eqs. (\ref{eq:8}), (\ref{eq:11}) give the possibility to
find the proper time of chemical freeze-out $\ \tau _{ch}$ if one knows the
''initial conditions'' at thermal freeze-out $\tau _{th}$: temperature $T$,
transverse scale parameter $R_{T}$ and velocity at the Gaussian border $%
\stackrel{.}{R}_{T}$. These values can be found from analysis of $p_{T}$%
-spectra and HBT correlations. Then solving (\ref{eq:8}), (\ref{eq:11})
numerically with initial conditions at thermal freeze-out, we find $T(\tau )$%
, and then find $\tau _{ch}$ from the equation $T(\tau _{ch})=T_{ch}$. It
gives the possibility to find the effective volume (\ref{eq:0.2}) at
chemical freeze-out, $V_{eff}^{ch}$. Further, using the absolute value of
the rapidity density for some particles, e.g., for pions, we can find the
common chemical potential $\mu _{ch}$ and, so, the common fugacity - the
factor that changes the energy and particle number densities as compared to
chemically equilibrated ideal gas. Finally all densities at chemical
freeze-out can be determined. Using the Eqs. (\ref{eq:5.05}), (\ref{eq:5.07}%
) one can also obtain the transverse velocity $v(r,\tau )$ at chemical
freeze-out.

\section{Reconstruction of Thermal Freeze-out Stage}

To reconstruct the thermal freeze-out conditions, we assume that the
freeze-out happens in a rather narrow time interval at approximately
constant temperature $T_{th}$. It corresponds in our solution to some proper
time $\tau _{th}\approx const$. Of course, the different particle species
can have different freeze-out (proper) times, so the time $\tau _{th}$\ is
just some mean value, when majority of strong interacting particles (pions,
protons, etc.) leave the system. Such an approximation gives no big
influence on the reconstruction of hadronization stage because, as we will
show, the deviations of different freeze-out times from that mean value for
majority of particles are relatively small. As for hyperons, like $\Omega $,
that can escape from the system essentially earlier, it also does influent a
little on hydrodynamic evolution because of a relatively small numbers of
those particles.

Within our fitting procedure we do not use temporal widths of thermal
freeze-out hypersurface because of two reasons.

First, such a width is introduced usually just as a common pure temporal
factor to the local equilibrium functions. It contradicts to natural picture
of emission, where the space-time correlations have to be presented:
particles from a periphery escape more easily than from the center (see
discussion about it in Ref. \cite{Akkelin-1}). The second point is that such
a temporal smoothing of locally equilibrium emission function is incorrect,
at least, if one based on Boltzmann equation (BE) \cite{Akkelin}. For
example, in order to calculate spectra in the case of exact locally
equilibrium solution of BE, one can use either Cooper-Frye prescription \cite
{Cooper} (sudden freeze-out) or, alternatively and equivalently, continuous
emission with{\bf \ }{\it non-equilibrium}{\bf {\it \ }}form of emission
function as it is shown in Ref. \cite{Akkelin}.

We choose the transverse profile factor in Gaussian form, $\rho (\tau
,r)=\exp (-r^{2}/2R_{T}^{2}(\tau ))$, as we noted before. Then we have for
the effective volume (\ref{eq:0.2}):
\begin{equation}
V_{eff}\approx 2\pi \tau R_{T}^{2}(\tau ).  \label{veff-a}
\end{equation}
To extract the temperature and transverse flows at thermal freeze-out we fit
in our model the experimental spectral slopes for different particle species.

The transverse spectra for direct particles can be expressed through the
Wigner function:
\begin{equation}
\frac{d^{2}N}{2\pi m_{T}dm_{T}dy}=\int d\sigma _{\mu }^{th}(x)p^{\mu
}f\left( \frac{p_{\mu }u^{\mu }(x)}{T_{th}},\frac{\mu _{i,th}}{T_{th}}%
\right) \rho (x).  \label{mt-spectra}
\end{equation}
At a region of small $p_{T}$ \ the contribution from the decay of resonances
has a peak that results in a significant change of the slope of transverse
spectra. At large $p_{T}$ , the relative contributions of a resonance decays
are distributed much more uniformly in $p_{T}$, especially if transverse
flow takes place, so, while spectra are shifted to higher values due to
resonance decays, the spectra slopes do not change essentially and the
slopes of the spectra are approximately the same as for direct particles
\cite{Stachel,Heinz-2}. Below we neglect the contribution of resonance
decays into the{\bf \ }{\it spectra slopes} in that region. At very high
energies $p_{T}>2$ GeV particles from the hard partonic processes at the
early collision stage may contribute to the spectra. We are fitting the
transverse spectra according to Eq. (\ref{mt-spectra}) in the $m_{T}-$%
interval between 0.3 and 2 GeV/c for pions, kaons and protons. For hyperons
we use an interval up to 3 GeV/c.

The other important point is how our results are influenced by the velocity
profile. The linearity of the transverse velocity in $r$ (\ref{eq:5.05}) is
associated with the non-relativistic approach for transverse flow (\ref
{nonrelat1}) and is valid for $r\lesssim R_{T}$ . Note that the integration
in Eq. (\ref{mt-spectra}) is carried out over all $r$-region requiring to
use correct velocity distributions when $v(\tau ,r)<1$ in the whole region $%
r<\infty $. So we have to extrapolate relativistically the velocity
distribution in the $r$-region of applicability of our hydrodynamic
solution, where $v^{2}(\tau ,r)\ll 1,$ to larger $r$ . For this aim we use
the results of Refs. \cite{Akkelin-1,Sinyukov-3}$\ $where two ''extreme''
relativistic extrapolations of a linear transverse velocity distribution
from non-relativistic region to relativistic one was used:
\begin{equation}
\ v(\tau _{th},r)=\tanh (rv^{\prime }(\tau _{th},0))  \label{hard}
\end{equation}
and
\begin{equation}
\ \ v(\tau _{th},r)=rv^{\prime }(\tau _{th},0)/\sqrt{1+(rv^{\prime }(\tau
_{th},0))^{2}},  \label{soft}
\end{equation}
where $v(\tau _{th},r)\approx rv^{\prime }(\tau _{th},0)$ for small
transverse velocity $v(\tau _{th},r).$ The extrapolation (\ref{hard})
corresponds to ''hard'' flow, while the extrapolation (\ref{soft}) - to
''soft'' one \cite{Akkelin-1,Sinyukov-3}. We have used both of them to fit
spectra and found that the deviation in extracted temperature $T_{th}$ as
well as in $v^{\prime }(\tau _{th},0)$ is about $5$ percent. It means also
that for SPS energies the relativistic tails of transverse velocity profile
have no essential influence on the spectra in above mentioned momentum
region, the transverse expansion, therefore, is basically non-relativistic.
Thus we reproduce the plots of spectra for only one velocity profile
corresponding to the ''hard'' flow, see also discussion in Refs. \cite
{Akkelin-1,Sinyukov-3}. $\ ${\bf \ }

There are two fitting parameters in the analysis of the slopes of transverse
spectra according to Eq.(\ref{mt-spectra}): $T_{th}$ and $\alpha
_{th}=(v^{\prime }(\tau _{th},0)R_{T})^{-2}$ , where {$1/\sqrt{{\alpha }}$
is the intensity of transverse flow \cite{Akkelin-1,Sinyukov-3,Xu}}

\begin{equation}
\sqrt{\alpha }=\frac{(v_{T}^{\prime }(\tau ,0))^{-1}}{R_{T}}=\frac{%
\mbox{hydrodynamical length}}{\mbox{transverse radius}}.  \label{alph-def}
\end{equation}
One can see that {$1/\sqrt{{\alpha }}$ is the hydrodynamic velocity at the
Gaussian ''boundary'' of the system, }
\begin{equation}
v_{R_{T}}(\tau )\equiv {v(\tau ,r=R_{T}(\tau ))=1/\sqrt{{\alpha }},}
\label{al}
\end{equation}
{\ within a linear approximation for the velocity profile, and according to
Eqs. (\ref{eq:5.05}), (\ref{eq:5.07}) }
\begin{equation}
{1/\sqrt{{\alpha }}=\ }\stackrel{.}{R}_{T}.  \label{alpha-def-new}
\end{equation}
For very intense relativistic transverse flow, $\alpha \rightarrow 0$. On
the other hand for $\alpha \gg 1$ we have non-relativistic transverse flow.
The results of the fit are presented at Fig. 1 and Fig. 2.

\newpage

\begin{figure}
\centerline{\hskip -0.cm%\hspace*{1.0cm}
\epsfig{width=15.cm,clip=1,figure=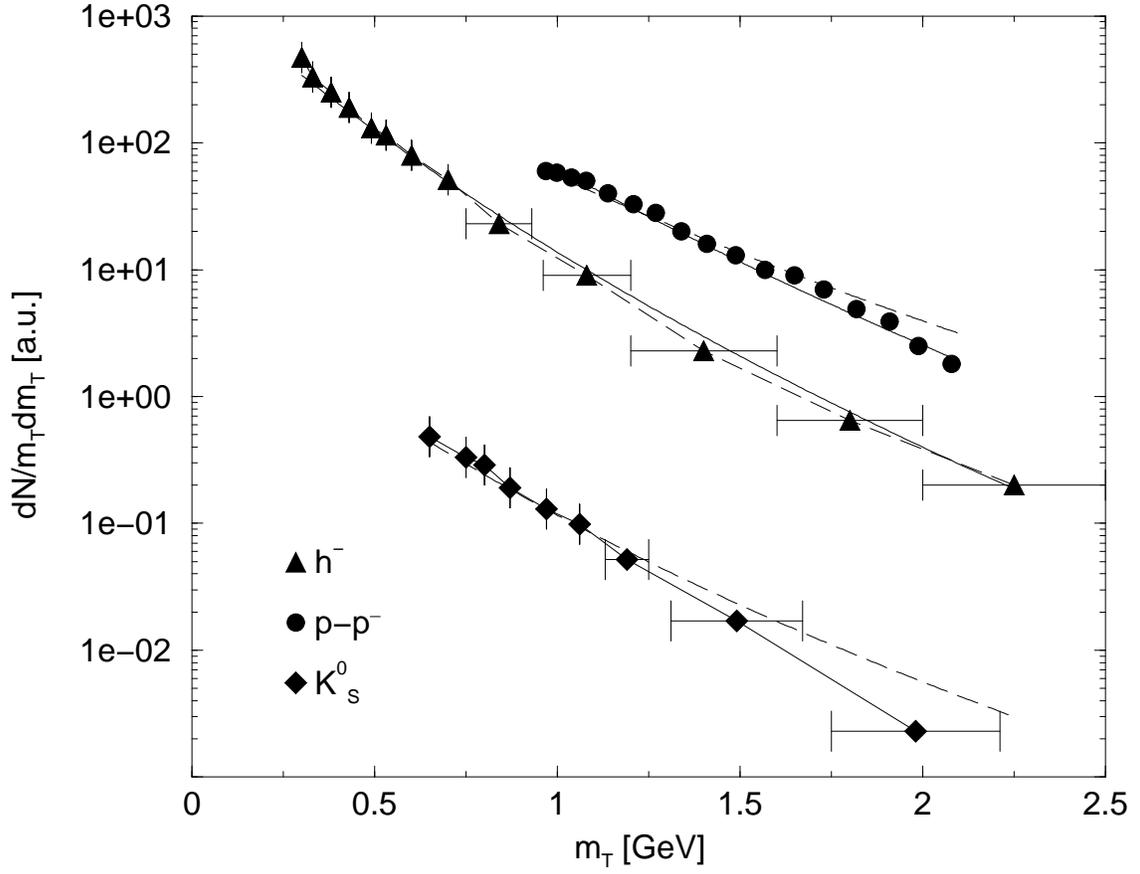} } \vspace*{-0.2cm}

\medskip
\caption{ The one-particle transverse mass spectrum for
$p-\overline{p}$ \protect\cite{NA49-4}, $h^{-}$ and $K_{S}^{0}$
\protect\cite{WA97-1}. Solid
line corresponds to $T_{th}=135$ MeV and $\alpha_{th} =7$, dashed line: $T_{th}=100$ MeV and $%
\alpha_{th} =4$. The overall normalization is arbitrary. }
\end{figure}

\newpage

\begin{figure}
\centerline{\hskip -0.cm%\hspace*{1.0cm}
\epsfig{width=15.cm,clip=1,figure=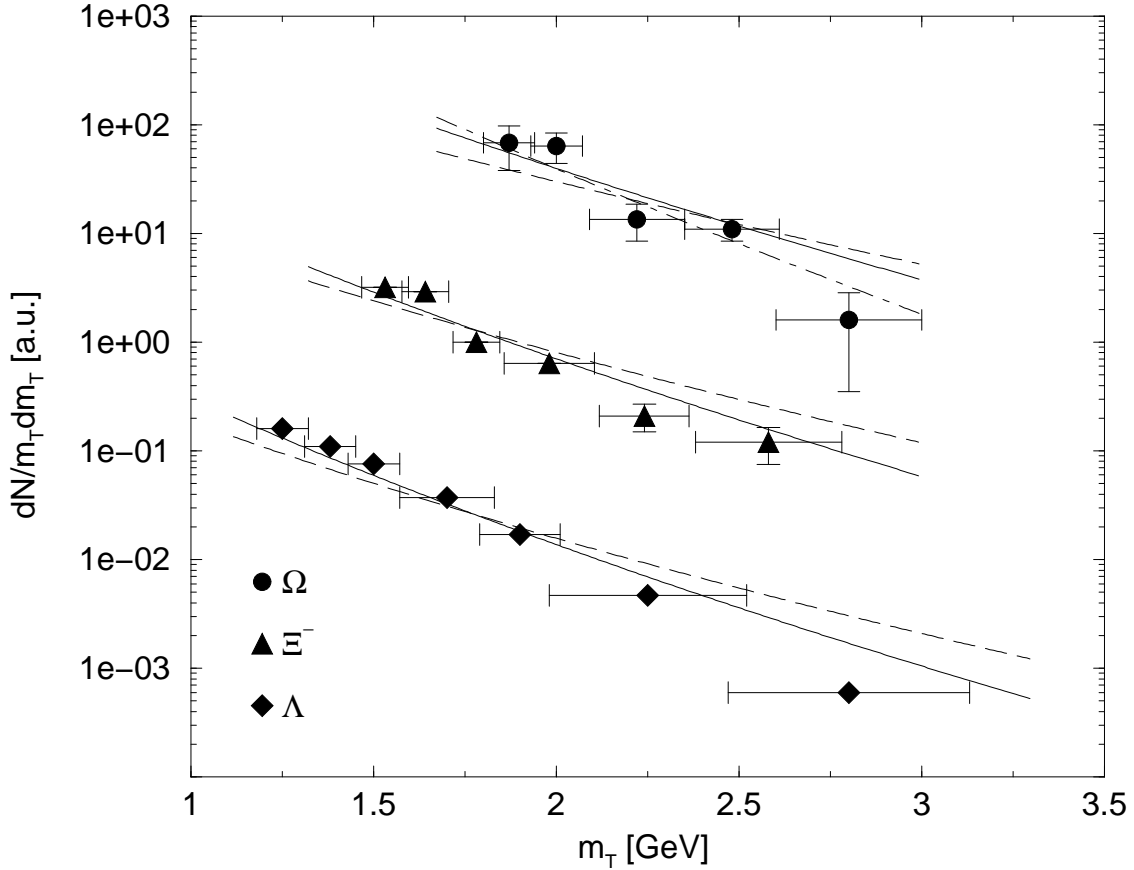} } \vspace*{-0.2cm}

\medskip
\caption{ The one-particle transverse mass spectrum for $\Omega=\Omega^{-}+%
\Omega^{+}$, $\Xi^{-}$ and $\Lambda$ \protect\cite{WA97-1}. Solid line: $%
T_{th}=135$ MeV and $\alpha_{th}=7$, dashed line: $T_{th}=100$ MeV
and $ \alpha_{th}=4$, dash-dotted line correspond to chemical
freeze-out parameters: $T_{ch}=163.37$ MeV and $\alpha_{ch}=10.8$.
The overall normalization is arbitrary. }
\end{figure}

As one can see from Fig. 1 the pion spectra can be described with good
accuracy in two regimes: relatively low temperature of thermal freeze-out, $%
T_{th}=100$ MeV \ and relatively intensive transverse flow, $\alpha _{th}=4$%
, on the one hand and higher temperature $T_{th}=135$ MeV and
non-relativistic transverse flow (within the typical size of the system at
freeze-out $R_{T}$) , $\alpha _{th}=7$, on the other hand. While the latter
parameterization describes well the spectra of all particles, except,
possible, the baryon $\Omega $, \ the former cannot reproduce the data on
spectral slopes except for pions (see Figs. 1, 2).

It is worth to note that the transverse spectra for $\Omega $ \ could be
described rather well, if one assumes that it escapes from the system just
after hadronization with corresponding chemical freeze-out parameters: $%
T_{ch}=163.37$ MeV and $\alpha _{ch\text{ }}=10.8$ , the latter is evaluated
from the hydrodynamic equations (see next section) with ''initial'' values $%
T_{th}=135$ MeV \ and $\alpha _{th}=7$ at thermal freeze-out. The
satisfactory result for the $\Omega $ -transverse spectra with chemical
freeze-out parameters, presented in Fig. 2, justifies the assumption of
earlier freeze-out for $\Omega $ \ that could happen just after
hadronization due to small cross-sections of these particles.

Finally we fix the thermal freeze-out parameters $T_{th}=135$ MeV \ and $%
\alpha _{th}=7$ according to the best fit for the set of transverse spectral
slopes for pions, kaons, protons, $\Lambda ,$ $\Xi $\ \ and $\Omega $. The
reason to use another, low temperature parameterization \footnote{%
Note here, to avoid the misunderstanding, that the thermal freeze-out
parameters $T=100$~MeV and $\alpha =4$ do not correspond to the our solution
of hydrodynamic equations with the initial conditions at $T=135$~MeV\ and $%
\alpha =7$.}, was that relatively large transverse flow gives the
possibility to describe satisfactorily the decrease of the observed pion
interferometry radii with increasing of $p_{T}$ \cite{Wiedemann}. As a
consequence, the temperature of thermal freeze-out becomes low, about $100$
MeV. Such interplay between flow and temperature is easy to understand
qualitatively from a simple approximation for spectral slopes:
\begin{equation}
T_{i,eff}\approx T_{th}+m_{i}/\alpha _{th},  \label{T-T}
\end{equation}
that is valid for heavy mass particles and non-relativistic flow \cite{Xu}.
However, as one can see from Fig. 1 and Fig. 2, such a low temperature
parameterization fails to reproduce the spectra slopes simultaneously for
many particle species, as we demonstrate for $p,K,\Lambda ,$ $\Xi $\ \ and $%
\Omega $ particles.

To discuss the problem, we would like to draw attention to resonance decay
as an important factor that could influence the behaviour of the
interferometry radii in $p_{T\text{ }}$, a factor that was not yet taken
into account quantitatively. It is rather difficult because one needs to
take into account of a large number of resonances. We note that the wide
spread opinion that due to relatively small temperatures in hadron resonance
gas one can use only resonances with low masses is not quite correct and, at
least, needs verification, because the number of resonance states at a fixed
mass interval increase quickly with mass. \

Generally, the contribution of resonance decays into the final particle
correlation function is a rather complicated effect. Namely, the resulting
resonance contribution to the interferometry radii depends mostly on the
following: statistical weight and mass distribution of resonances in the
rest frame of fluid element, the intensity of flow, lifetime of resonance,
feeding from middle-lived resonances, such as $\omega $, and the kinematic
''window'' for pions coming from resonance decay. The important factor is
also the length of the region of homogeneity \cite{Sinyukov-2}, that is
effective size of fluid element emitting the resonance in some $p_{T\text{ \
}}$region, for developed flows it is rather less than the corresponding
values for pions due heavier resonance masses. All these factors vary for
different resonances and we cannot {\it a priory} exclude or easily estimate
the influence of resonance decays on the interferometry radii at CERN SPS
A+A collisions. Qualitatively, the steep decrease of the observed
interferometry radii on $p_{T}$ could be due to an interplay between two
factors: transverse expansion of the hydrodynamic tube and contribution to
the radii from resonance decays. For relativistic transverse flow the
contribution of low-mass resonance decays to the interferometry radii is
negligible \cite{Sinyukov-1},\cite{Heinz-2} and the decrease is determined
mostly by flow. If the transverse expansion is not very intensive, the
decrease of the interferometry radii on $p_{T}$ is the result of both
effects: flow and decays of resonances \cite{Bolz}. Therefore, in our
opinion, the detailed evaluation of the resonance contribution to
interferometry radii in hydrodynamic approach needs a serious study, taking
into account the contribution of high mass resonances. In this paper we
analyze the interferometry radii for pions, kaons and protons at maximal
measured $p_{T}$ region only, in order to minimize the influence of
resonance contribution. Here we are{\bf \ }based on known results \cite
{Heinz-2,Bolz} of hydrodynamic approach and our preliminary studies with
relatively small number of resonances \cite{Sinyukov-1}. According to these
results, the resonance decays lead to an increase of the interferometry
radii at small $p_{T}<p_{T0}$\ ( $p_{T0}\approx 0.3\div 0.4$\ GeV for pions)
and to a vanishing of resonance contributions into the radii out of this
region. The latter means that the complete interferometry radii and
interferometry radii for ''direct'' particles (calculated neglecting
resonance decays) are rather close to each other at $p_{T}>p_{T0}$.

Finally, we would like to note that the aim of the paper is not the
description of all spectra and correlations in the whole momentum region but
the reconstruction of hadronization conditions. For this aim it is enough to
use the slopes of spectra only and interferometry radii in some momentum
interval. As for the description of spectra and correlations in whole
momentum region, it is a separate problem that demands a detailed analysis
within our hydrodynamic model of decays of huge number of resonances, and
taking into account the back reactions to these decays at all stages of
evolution of hadronic system (to calculate the number of resonances which
are survived till the thermal freeze-out).

Our calculation of Bose-Einstein and Fermi-Dirac correlation function,
\begin{equation}
C^{ij}(p,q)=\;p_{1}^{0}p_{2}^{0}\frac{d^{6}N_{ij}}{d^{3}{\bf p}_{1}d^{3}{\bf %
p}_{2}}/\left( p_{1}^{0}\frac{d^{3}N_{i}}{d^{3}{\bf p}_{1}}p_{2}^{0}\frac{%
d^{3}N_{j}}{d^{3}{\bf p}_{2}}\right) ,  \label{corr-def}
\end{equation}
is based on the distribution function $f(p_{\mu }u^{\mu }(r,\eta
)/T_{th}(\eta ),\mu _{i,th}(\eta )/T_{ch}(\eta ))$ in Boltzmann
approximation at the hypersurface of thermal freeze-out. For $\pi ^{-}\pi
^{-}$ correlation function we have, e.g.,

\begin{equation}
C(p,q)=\;1+\lambda \frac{\left| \int d\sigma _{\mu }^{th}(x)p^{\mu }f\left(
\frac{p_{\mu }u^{\mu }(x)}{T_{th}},\frac{\mu _{i,th}}{T_{th}}\right) \rho
(x)e^{iqx}\right| ^{2}}{\left( \int d\sigma _{\mu }^{th}(x)p_{1}^{\mu
}f\left( \frac{p_{1\mu }u^{\mu }(x)}{T_{th}},\frac{\mu _{i,th}}{T_{th}}%
\right) \rho (x)\right) \left( \int d\sigma _{\mu }^{th}(x)p_{2}^{\mu
}f\left( \frac{p_{2\mu }u^{\mu }(x)}{T_{th}},\frac{\mu _{i,th}}{T_{th}}%
\right) \rho (x)\right) }  \label{corr-pi}
\end{equation}
where $p=(p_{1}+p_{2})/2,$ $q=p_{1}-p_{2}$ and $\lambda $ is
phenomenological parameter describing the experimental suppression of the
correlation function due to, e.g., contribution of long-lived resonances. We
are fitting by $\chi ^{2}$ -method the correlation function with fixed
parameters $T_{th}=135$ MeV \ and $\alpha _{th}=7$ to Gaussian approximation
of experimental data: $C(p,q_{i},0,0)=1+\lambda \exp
[-q_{i}^{2}R_{i}^{2}(p)] $ in Bertsch-Pratt parameterization. Then $\ i=l,$ $%
s $, $o$ $\ $correspond to longitudinal ($l$) and two transverse $(t)$:
sideward $(s)$ and outward $(o)$ directions.

Our results for proper time $\tau _{th}$ and geometrical Gaussian radius $%
R_{T}$ of thermal freeze-out are presented in Table 3 in the last two
columns. The experimental interferometry radii $R_{l\text{ }}$ and $R_{s%
\text{ }}$for pion, kaon and proton pairs measured in LCMS near mid-rapidity
are taken at maximum possible $p_{T}$ . We assume, in agreement with
experimental data for pion distribution in rapidity, that\ the factor $\rho
(\tau ,r,\eta )$ does not vary strongly in rapidity within one unit of
rapidity near $\eta =0.$

\begin{center}
TABLE 3. The characteristics of the thermal freeze-out in Pb+Pb SPS
collisions. \footnote{%
Two-dimensional fits for kaon and proton correlation functions in \cite
{NA44-kaons} and \cite{Conin} were used .}

\begin{tabular}{|l|l|l|l|l|l|}
\hline
& $p_{T}$ GeV & $R_{l}$ fm & $R_{s}$ fm & $R_{T}$ fm & $\tau _{th}$ fm \\
\hline
$\pi ^{-}\pi ^{-}$ \cite{Ganz} & $0.5$ & $4.6$ & $4.25$ & $5.3$ & $8.2$ \\
\hline
$K^{+}$ $K^{+}$\cite{NA44-kaons} & $0.91$ & $3.2$ & $3.59$ & $5.1$ & $8.9$
\\ \hline
$pp$ \cite{Conin} & $1.073$ & $2.9$ & $3.4$ & $5.2$ & $9.5$ \\ \hline
\end{tabular}
\end{center}

Let us note that the proper time of thermal freeze-out $\tau _{th}$ and
transverse radius $R_{T}$ of system is slightly different for different pair
correlation functions. To reconstruct the chemical freeze-out we will use
the average values of the above parameters. In the Table 4 we compare our
results ($1$st fit) with the results of other authors. Temperature $T_{th}$,
proper time $\tau _{th}$, transverse Gaussian radius of the system $%
R_{T}(\tau _{th})$ and transverse velocity at Gaussian ''boundary'' of the
system, $v_{R_{T}}(\tau _{th})${$=1/\sqrt{\alpha _{th}}$, }at thermal
freeze-out\ for fits $2-4$ are taken from Refs.\cite{Wiedemann}, and for fit
$5$ from Ref. \cite{Ster}. Note, that in our model of post-hadronization
hydrodynamic evolution the transverse velocity at Gaussian ''boundary'' of
the system coincides with proper time derivative of transverse Gaussian
radius, see Eq. (\ref{alpha-def-new}). In the columns $4,5$ we present our
results for the values of chemical potential of direct pions $\mu _{\pi ,th}$
and particle number density (before resonance decay) $\overline{n}%
_{th}=n(\tau _{th},r=0,\eta =0)$ at the thermal freeze-out in the very
center of the system, which were calculated by the method described in the
next section for different parameters $T_{th}$, $\tau _{th}$, $R_{T}$ $(\tau
_{th})$, $v_{R_{T}}(\tau _{th})${\ } (fits $1-5$).

\bigskip

TABLE 4. The characteristics of the thermal freeze-out in Pb+Pb SPS
collisions.

\begin{center}
\begin{tabular}{|l|l|l|l|l|l|l|}
\hline
Fit & $T_{th}$ MeV & $\tau _{th}$ fm & $R_{T}(\tau _{th})$ fm & $%
v_{R_{T}}(\tau _{th})$ & $\overline{n}_{th}$ 1/fm$^{3}$ & $\mu _{\pi ,th}$
MeV \\ \hline
$1$ & $135$ & $8.9$ & $5.2$ & $0.378$ & $0.274$ & $74.3$ \\ \hline
$2$ & $100$ & $7.8$ & $6.5$ & $0.424$ & $0.199$ & $128.5$ \\ \hline
$3$ & $120$ & $6.6$ & $5.9$ & $0.339$ & $0.286$ & $119.7$ \\ \hline
$4$ & $160$ & $5.5$ & $5.6$ & $0.247$ & $0.381$ & $50.0$ \\ \hline
$5$ & $139$ & $5.9$ & $7.1$ & $0.55$ & $0.221$ & $32.9$ \\ \hline
\end{tabular}
\end{center}

\bigskip

At the first sight, the particle number density at the thermal freeze-out $%
\overline{n}_{th}=0.274$ 1/fm$^{3}$, which is obtained in our analysis, is
too high, formally, the mean free path for pions, as we will demonstrate
below, is rather small as compare with transverse ''size'' of the system, $%
R_{T}(\tau _{th})$. For static systems such a result could indicate the
absence of \ the freeze-out of momentum spectra at this temperature.
However, if the system expands intensively, the comparison of a rate of
expansion with the rate of collisions plays the dominant role for freeze-out
criterium \cite{Bondorf,Lee,Navarra,Ekkard}, \cite{Shuryak}, and is more
important for estimation of freeze-out parameters than the comparison of
mean free path with geometrical Gaussian ''size'' of the system. Namely, for
the system that we are interested in, local thermal equilibrium and ideal
fluid hydrodynamic behaviour can be maintained as long as the collision rate
among the particles is much faster than rate of expansion. Since densities
drop out during 3-dim expansion rather intensively, the collision rate
decreases rapidly and at some moment system falls out of equilibrium. As a
result, system decouples and the freeze-out happens. Then, to describe
momentum spectra we use the standard Cooper-Frye freeze-out prescription
\cite{Cooper}, that assumes validity of the ideal hydrodynamics up to a
''sharp'' 3-dim freeze-out hypersurface, whereas outside of this surface the
particles are supposed to move freely. It is worth to note, that Cooper-Frye
prescription can be applied only at densities where the matter is (locally)
in thermal equilibrium, and therefore {\it at the thermal freeze-out the
rate of expansion should be comparable with the rate of collisions but
smaller than it}. Let us check, whether the densities, that correspond to
our fitting parameters, Table $4$ (fit $1$), meet to such a freeze-out
criterium.

The inverse expansion rate is the collective expansion time scale \cite
{Ekkard}:
\begin{equation}
\tau _{\exp }=-\left( \frac{1}{n}\frac{\partial n}{\partial t^{\ast }}%
\right) ^{-1}=-\left( \frac{1}{n}u^{\mu }\partial _{\mu }n\right) ^{-1}=%
\frac{1}{\partial _{\mu }u^{\mu }},  \label{expansion-time}
\end{equation}
where $t^{\ast }$ is proper time in the local rest system of fluid and the
last equality follows from the conservation of particle number density
currents (\ref{eq:3.2}). Using for 4-velocities Eq. (\ref{transv}), one can
easily get that
\begin{equation}
\partial _{\mu }u^{\mu }=\frac{\partial }{\partial \tau }\gamma +\frac{%
\partial }{\partial r}(\gamma v)+\gamma (\frac{1}{\tau }+\frac{v}{r}).
\label{vel-grad1}
\end{equation}
Taking into account expression for transverse velocity (\ref{eq:5.05}), (\ref
{eq:5.07}), we get finally
\begin{equation}
\tau _{\exp }(\tau )\approx \left( \frac{1}{\tau }+\frac{2\stackrel{.}{R}_{T}%
}{R_{T}}\right) ^{-1}.  \label{vel-grad2}
\end{equation}
Using our parameters at\ the thermal freeze-out from Table $4$ (fit $1$) we
obtain that collective expansion time at thermal freeze-out $\tau _{\exp
}(\tau _{th})\approx 3.88$ fm. Note, that $\tau _{\exp }(\tau
_{th})<R_{T}(\tau _{th})=5.2$ fm.

The inverse scattering rate of particle species $i$ is the mean time between
scattering events for particle $i$, $\tau _{scat}^{(i)}$:
\begin{equation}
\tau _{scat}^{(i)}\approx \frac{1}{\sum \left\langle v_{ij}\sigma
_{ij}\right\rangle n_{j}},  \label{scattering-time}
\end{equation}
where $v_{ij}$ is the relative velocity between the scattering particles and
$\sigma _{ij}$ is the total cross section between particles $i$ and $j$, the
sharp brackets mean an average over the local thermal distributions. This
time is determined by the densities of all particles with which particle $i$
can scatter, and the corresponding scattering cross sections. Let us
estimate the mean time between scatterings for pions, $\tau _{scat}^{(\pi )}$%
. First, note that $\tau _{scat}^{(i)}>\lambda ^{(i)}$, where $\lambda
^{(i)} $ is mean free path for particle species $i$:
\begin{equation}
\lambda ^{(i)}\approx \frac{1}{\sum \left\langle \sigma _{ij}\right\rangle
n_{j}},  \label{mean-free}
\end{equation}
so $\lambda _{scat}^{(i)}$ represents the lower limit for $\tau
_{scat}^{(i)} $. Let us estimate the pion mean free path in the very center
of the system, $\lambda ^{(\pi )}(\tau _{th},r=0)$, using the following
approximation:
\begin{equation}
\lambda ^{(\pi )}(\tau _{th},0)\approx \frac{1}{\sigma _{\pi p}n_{B}+\sigma
_{\pi \pi }n_{M}}  \label{mean-ap}
\end{equation}
where $\sigma _{\pi p}=65$ mb is supposed to be total cross section for
pion-proton scattering, and we assume the same cross-section for all
non-strange baryons,\ whereas $\sigma _{\pi \pi }=10$ mb is total cross
section for pion-pion scattering \ and we assume the same cross-section for
all non-strange mesons. The densities $n_{B}=0.044$ 1/fm$^{3}$ and $%
n_{M}=0.159$ 1/fm$^{3}$ are the densities of non-strange baryons and
non-strange mesons respectively at $\tau =\tau _{th}$ and $r=0$, the
densities were calculated by the method described in the next section. The
contribution of strange particles to pion mean free path is neglected in (%
\ref{mean-ap}). Then $\lambda ^{(\pi )}(\tau _{th},0)\approx 2.26$ fm and
this value is essentially smaller than $R_{T}(\tau _{th})=5.2$ fm, while the
mean time between scattering events for pions $\tau _{scat}^{(\pi )}>\lambda
^{(\pi )}$ and, therefore, is less but comparable with the collective
expansion time $\tau _{\exp }(\tau _{th})\approx 3.88$ fm. So, one can
conclude that our fitting parameters correspond to freeze-out criterium.
Note, that the pion mean free path at the Gaussian ''boundary'' of the
system $\lambda ^{(\pi )}(\tau _{th},R_{T})\approx 3.73$ fm, and, thus, the
densities, that are represented in Table $4$ (fit $1$), are just at the
border where ideal hydrodynamics still can be applied. Lower values for
densities, on our opinion, are not appropriate to apply the Cooper-Frye
prescription within hydrodynamics.

Of special interest is the average transverse velocity, that depends on the
equation of state during the thermalized matter evolution after collision.
At central slice of hydrodynamic tube the mean transverse hydrodynamic
velocity $\langle v\rangle \equiv $ $\left\langle \left| {\bf v}_{T}(\eta
=0)\right| \right\rangle $ and mean square of transverse hydrodynamic
velocity $\langle v^{2}\rangle $ are defined as follows:
\begin{equation}
\left\langle v^{n}\right\rangle =\frac{\int \frac{d^{3}p}{p^{0}}\int \frac{%
d\sigma _{\mu }}{d\eta }p^{\mu }v^{n}(x)f(x,p)}{\int \frac{d^{3}p}{p^{0}}%
\int \frac{d\sigma _{\mu }}{d\eta }p^{\mu }f(x,p)}=\frac{\int \frac{d\sigma
_{\mu }}{d\eta }u^{\mu }(x)\rho (x)v^{n}(x)}{\int \frac{d\sigma _{\mu }}{%
d\eta }u^{\mu }(x)\rho (x)},\quad n=1,2.  \label{vt}
\end{equation}
In the first approximation in $1/\alpha $ expansion one can get for $\alpha
\gg 1.$
\begin{equation}
\left\langle v\right\rangle \approx \sqrt{\frac{\pi }{2\alpha }},\quad
\left\langle v^{2}\right\rangle \approx \frac{2}{\alpha }.  \label{v-app}
\end{equation}
This approximation corresponds to non-relativistic flow and is
model-independent (the result does not depend on relativistic extrapolation
of the velocity profile, see discussion above). Then, at thermal freeze-out (%
$\alpha _{th}=7)$ we get for the average transverse velocity of the central (%
$\eta =0)$ slice of the hydrodynamic tube:$\;\left\langle v\right\rangle
_{th}\approx 0.47$.

Note, that using (\ref{T-T}), one can get for cylindrically symmetric
expansion the expression for transverse spectral slopes
\begin{equation}
T_{i,eff}\approx T_{th}+\frac{1}{2}m_{i}\left\langle v_{th}^{2}\right\rangle
=T_{th}+\frac{2}{\pi }m_{i}\left\langle v_{th}\right\rangle ^{2}.
\label{slopes}
\end{equation}
Right-hand side in the first equality is approximately the ratio of average
transverse energy of particles species ''$i$'' in central slice, $%
\left\langle \frac{dE_{i,T}}{d\eta }\right\rangle \approx \frac{dN_{i}}{%
d\eta }T_{th}+\frac{dN_{i}}{d\eta }m_{i}\left\langle v_{th}^{2}\right\rangle
/2$ \ to total number of particles of such species in central slice, $\frac{%
dN_{i}}{d\eta }$, at thermal freeze-out.

\section{Hydrodynamic reconstruction of chemical freeze-out stage}

Based on the hydrodynamic solutions found in Sec. IV and ``initial
conditions'' at thermal freeze-out considered in Sec. V, we can restore the
space-time extension of hadronization stage and find the particle and energy
densities. We use the solution of hydrodynamic equation in the center of the
system, where the majority of particles are produced and where
non-relativistic approximation for transverse velocity is correct.{\bf \ }In
Fig. 3 we demonstrate the cooling down of HG as it depends on proper time $%
\tau $ for fit 1. The ``initial condition'' for the hydrodynamic solution is
the temperature of thermal freeze-out $T_{th}=135$ MeV at corresponding time
$\tau _{th}=8.9$ fm/c which were found from the spectra and correlations in
the previous section. The ''initial'' velocity ( at $\tau _{th}$) at the
Gaussian ''boundary'' of the system $v_{R_{T}}(\tau _{th})\approx 0.38$
justifies the applicability of our approach, namely, \ non-relativistic
transverse expansion (\ref{nonrelat1}) in central domain: $%
v_{R_{T}}^{2}(\tau _{th})=\stackrel{.}{R}_{T}^{2}(\tau _{th})\approx 0.14\ll
1$ and so the restriction (\ref{nonrelat3}) is also satisfied between
chemical and thermal freeze-out. The second condition (\ref{nonrelat2}) and
restriction (\ref{nonrelat4.4}), that follows from it, is also satisfied: $%
T_{ch}/(e(T_{ch})+T_{ch})\approx 0.141\ll 1$ and $T_{th}/(e(T_{th})+T_{th})%
\approx 0.127\ll 1$.

\newpage

\begin{figure}
\centerline{\hskip -0.cm%\hspace*{1.0cm}
\epsfig{width=15.cm,clip=1,figure=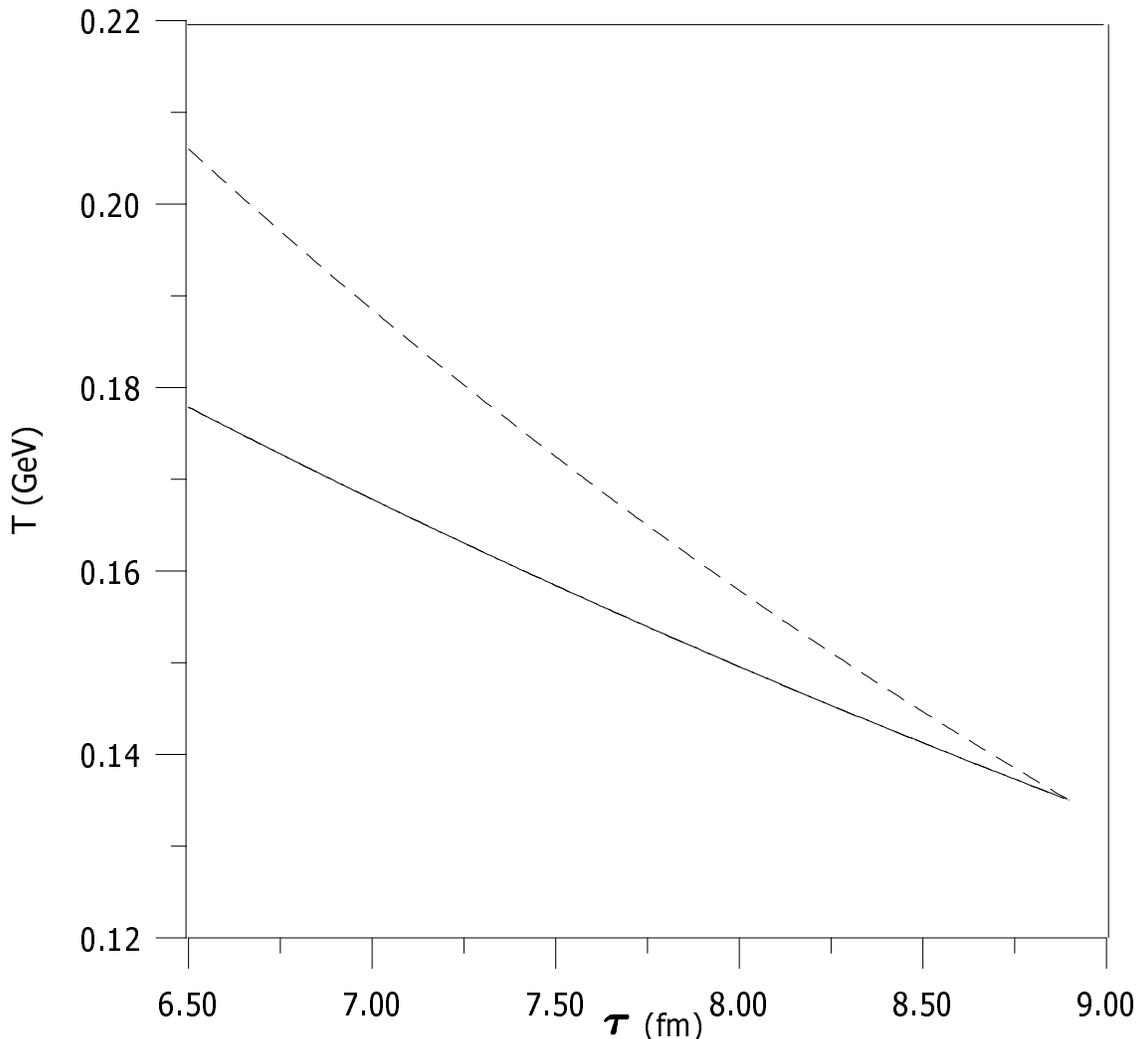} } \vspace*{-0.2cm}

\medskip
\caption{The temperature T (GeV) of the hadron resonance gas as
function of the proper time $\tau$ (fm/c) of its evolution. Solid
line corresponds to the semi-relativistic solution $T(\tau)$ of
Eqs. (\protect\ref {eq:8}) and (\protect\ref{eq:11}). Dashed line
describes $T(\tau)$ in
completely non-relativistic picture of evolution of Boltzmann gases, Eqs. (%
\protect\ref{eq:9}) and (\protect\ref{eq:13}), with the same
initial conditions. The effective mass of the mixture of
non-relativistic Boltzmann gases is $m=\sum \kappa
_{i}m_{i}\approx 0.662$ GeV. }
\end{figure}

One can restore the proper time of the chemical freeze-out from the
intercept of the curve $T(\tau )$ at the temperature of the chemical
freeze-out $T_{ch}\approx 163.37$ MeV, that was found from the analysis of
particle number ratios. Using the result, $\tau _{ch}=7.24$ fm/c, one can
calculate the effective volume $V_{eff}^{ch}\approx 2\pi \tau
_{ch}R_{T}^{2}(\tau _{ch})\approx 975$ fm$^{3}$ associated to unit of
rapidity at mid-rapidity, and then find common chemical potential $\mu _{ch}$
from pion multiplicities at mid-rapidity, $dN_{\pi ^{-}}/dy\approx 160$ \cite
{NA49-1}. For direct pions the common chemical potential $\mu _{ch}$ is
added to phenomenological one: $\mu _{\pi ,ch}=$ $\mu _{\pi }+\mu _{ch}$.
Then one can easily evaluate the energy and particle densities.

Before proceeding to numerical results, let us check the validity of our
basic approximations. The duration of hydrodynamic expansion of hadron gas, $%
\Delta \tau =8.9-7.24=1.66$\ fm/c, is smaller or comparable with the
relaxation time to chemical equilibrium in the system (see, e.g., \cite
{Song,Rapp}), that exclude the chemical equilibration during the evolution.\
The typical life-time of the short lived resonances, is, however, compared
with the period of hadron hydrodynamic stage $\Delta \tau =1.66$ fm/c,
therefore, such decays could take place at the background of chemically\
non-equilibrium expansion. The decays could result in an increase of number
of (quasi) stable particles during the expansion (without that all particle
numbers will conserve within concept of chemical freeze-out). On the other
hand, there are back reactions, that produce the resonances. These reactions
do not lead to chemical equilibrium, but just reduce effect of resonance
decays into the chemically non-equilibrium medium, resulting in an increase
of effective life-time of resonances.

At the same time, some residual effect of resonance decays during the
evolution could, anyway, take place. We ignore that effect since our
approximate calculation of resonance decays demonstrate two opposite
influences of it on a fluid evolution. (To do this estimate we replace the
discrete spectra of heavy resonances by the Hagedorn distribution). The
decay of heavier resonances with subsequent thermalization of decay products
leads to a heating of the system, while the decay of relatively light
resonances results in cooling. The reason is clear: the difference between
kinetic energy of decay products and kinetic energy of thermal motion of the
same particles (e.g. pions) is positive in the former case and negative in
the latter one. Hence, the corresponding effects are approximately
compensated and give no big influence on hydrodynamic evolution.

Our results are presented in Table 5. The first line corresponds to
parameters at thermal freeze-out which are represented in the first line of
Table 4. In Table 5 we present also our results for proper time and
densities at the hadronization stage based on the thermal freeze-out
parameters taken from other papers: fits 2-5 from Table 4.

From the results presented in Table 5 one can see that the ratio of energy
density to total particle density at chemical freeze-out reaches the value $%
\varepsilon _{ch}/n_{ch}=0.997$ GeV, in agreement with the phenomenological
observation in Ref. \cite{Cleymans}. The energy density at the hadronization
point $\overline{\varepsilon }_{ch}\approx 0.42$ GeV/fm$^{3}$ is close to
results of lattice calculation at the region of the phase transition and
about equal to the energy density inside a nucleon. Note that for
temperature and chemical potential found in \cite{P. Br.-M.-2} at the
chemical freeze-out, we get $\overline{\varepsilon }_{ch}\approx 0.54$ GeV/fm%
$^{3}$ and $\varepsilon _{ch}/n_{ch}\approx 1.16$ GeV, that is close to our
results. Using the values of concentrations of HG and temperatures at
chemical and thermal freeze-out, we get for conventional velocity of sound $%
c_{0}^{2}=p/\varepsilon =T/\sum \kappa _{i}e_{i}(T)$ the values $0.164$ and $%
0.145$ at chemical freeze-out and thermal freeze-out, respectively. Such
relatively small values are determined by the large average mass of
particles in hadron resonance gas, that is $0.662$ GeV for hadronic masses
below $2$ GeV.

In Table 4 we list also the total particle number density $\overline{n}_{th}=%
\overline{n}_{ch}V_{eff}^{ch}/V_{eff}^{th}$, and chemical potential for
direct pions $\mu _{\pi ,th}$ at thermal freeze-out. One can see that for
small temperature fits, $T_{th}=100\div 120$ MeV (fits 2 and 3) the chemical
potential of direct pions at the stage of system decay seems to be too large
and closed to condensation point, $\mu _{\pi ,th}=m_{\pi }$.

Using Eqs. (\ref{alpha-def-new}), (\ref{v-app}) and velocity at Gaussian
''boundary'' of the system at the chemical freeze-out $v_{R_{T}}(\tau _{ch})$%
{$=$}$\stackrel{.}{R}_{T}(\tau _{ch})$, we calculate the averaged velocity
at chemical freeze-out $\left\langle v\right\rangle _{ch}$ for fit $1$ and
find that $\left\langle v\right\rangle _{ch}\approx 0.383$.

\begin{center}
TABLE 5. The characteristics of the chemical freeze-out in Pb+Pb SPS
collisions.

\begin{tabular}{|l|l|l|l|l|l|l|}
\hline
Fit & $\tau _{ch}$ fm & $\mu _{ch}$ MeV & $R_{T}(\tau _{ch})$ fm & $%
v_{R_{T}}(\tau _{ch})${$=$}$\stackrel{.}{R}_{T}(\tau _{ch})$ & $\overline{%
\varepsilon }_{ch}$ GeV/fm$^{3}$ & $\overline{n}_{ch}$ 1/fm$^{3}$ \\ \hline
$1$ & $7.24$ & $-67$ & $4.63$ & $0.304$ & $0.420$ & $0.421$ \\ \hline
$2$ & $4.1$ & $-9.8$ & $5.18$ & $0.278$ & $0.598$ & $0.600$ \\ \hline
$3$ & $4.2$ & $-13.5$ & $5.19$ & $0.245$ & $0.583$ & $0.585$ \\ \hline
$4$ & $5.3$ & $-76.1$ & $5.55$ & $0.240$ & $0.398$ & $0.399$ \\ \hline
$5$ & $4.8$ & $-110.4$ & $6.53$ & $0.491$ & $0.323$ & $0.324$ \\ \hline
\end{tabular}
\end{center}

\section{Conclusions}

The reconstruction of the hadronization stage for CERN SPS Pb+Pb collisions
at 158A GeV/c is obtained within the hydrodynamic approach for the
post-hadronization stage. Because of incomplete stopping at SPS energy we
demonstrate that the analysis of particle ratios can be done within a
relatively narrow rapidity window near mid-rapidity, provided one takes a
non-uniformity in $\mu _{B}$ into account. We estimate that the
non-uniformity of baryochemical potential may result in small but noticeable
($\approx 10$ $\%$) contribution to the pion multiplicity at mid-rapidity
due to transport of pions from resonance decays from non-central rapidity
regions. Using\ an optimization procedure for the calculation of particle
number ratios near mid-rapidity, and also reconstructing the thermal
freeze-out stage from spectra and correlations for different particle
species, we find the physical conditions at the hadronization stage. The
proper time for chemical freeze-out is found to be too small, $\tau
_{ch}\approx 7.2$ fm/c to reach the chemical equilibrium in pure hadron gas,
or cascade ``event generator'', approach. We stress that our method does not
assume complete chemical equilibrium of Boltzmann gases at the
hadronization, and allow to estimate the deviation from it. We find that
distribution functions of hadrons at hadronization hypersurface include some
common factor, $\exp (\mu _{ch}/T_{ch})\approx 0.66$, that reduces
essentially particle numbers as compare to chemically equilibrium ones. The
energy density is found to be close to the results of lattice calculations
at the point of the phase transition. The ratio of energy density to total
particle number density is found to be about 1 GeV.

All those results indicate the existence of a pre-hadronic stage of
expansion before the chemical freeze-out and the comparison with lattice
calculations pointed to a QGP phase as most probable state of the matter
till $\tau \simeq $\ $7.2$\ fm/c in Pb+Pb 158 A GeV/c collisions. The
obtained characteristics of the hadronization give the possibility to make
the next step and reconstruct the pre-hadronic stage of evolution and, then,
to study the basic properties of \ QGP. \

\section*{Acknowledgments}

This work was supported by German-Ukrainian Grant No. 2M/141-2000, Ukrainian
- Hungarian Grant No. 2M/125-99 and French-Ukrainian CNRS Grant No. Project
8917.

\end{document}